\documentclass[aps,prc,12pt,nofootinbib,superscriptaddress,showkeys,english]{revtex4-1}
\usepackage[english]{babel}
\usepackage{graphicx}
\usepackage{color}
\usepackage{amsmath}
\usepackage{amssymb}
\usepackage{multirow}
\usepackage{color}

\begin{document}
\title{Finite temperature effects on magnetized Bose-Einstein condensate stars}
\author{G. Quintero Angulo}
\email{gretica.qa@gmail.com}
\affiliation{Facultad de F{\'i}sica, Universidad de la Habana,\\ San L{\'a}zaro y L, Vedado, La Habana 10400, Cuba}
\author{L. C. Su\'arez Gonz\'alez}
\email{lismarydelacaridad@gmail.com}
\affiliation{Instituto de Cibern\'{e}tica, Matem\'{a}tica y F\'{\i}sica (ICIMAF), \\
	Calle E esq a 15 Vedado 10400 La Habana Cuba}
\author{A. P\'erez Mart\'inez\footnote{On leave from Departamento de F\'isica Te\'orica, Instituto de Cibern\'etica Matem\'atica y F\'isica (ICIMAF), Cuba}}
\email{aurorapm1961@usal.es}
\affiliation{Departamento de F\'sica Fundamental, Universidad de Salamanca, Plaza de la Merced s/n 37008, Espa\~na}
\author{H. P\'erez Rojas}
\email{hugo@icimaf.cu}
\affiliation {Instituto de Cibern\'{e}tica, Matem\'{a}tica y F\'{\i}sica (ICIMAF), \\
 Calle E esq a 15 Vedado 10400 La Habana Cuba}

\date{\today}
\begin{abstract}
	
We study the role of temperature and magnetic field on the equation of state and macroscopic properties of Bose-Einstein condensate stars. These compact objects are composed of a condensed gas of interacting neutral vector bosons coupled to a uniform and constant magnetic field. We found that the main consequence of a finite temperature in the magnetized equations of state is to increase the inner pressure of the star. As a consequence, magnetized hot Bose-Einstein condensate stars are larger and heavier than their zero-temperature counterparts. However, the maximum masses obtained by the model remain almost unchanged, and the magnetic deformation of the star increases with the temperature. Besides, augmenting the temperature reduces the number of stable stars, an effect that the magnetic field enhances. The implications of our results for the star's evolution, compactness, redshift, and mass quadupolar moment are also analyzed.

\end{abstract}


\maketitle

\section{Introduction}
\label{sec1}

In the last decade, observational evidence associated with the theoretical adjustment of the cooling data of Cassiopeia A has come to reinforce the long-lasting supposition that neutron stars (NS) may mainly consist of a superfluid of neutrons \cite{Shternin2011,Page:2011yz,PageSC}. Considering the NS's typical inner conditions, paired neutrons are expected to be in an intermediate situation between the Bardeen-Copper-Schrieffer and the Bose-Einstein condensate (BEC) limits, prevailing the former description \cite{Gruber2014}.
However, assuming the pairs in the BEC limit led, some years ago, to Bose-Einstein condensate stars (BECS) models \cite{chavanis2012bose,chavanis2012bose,latifah2014bosons,Gruber2014,quintero2019self,quintero2019magnetized}. A BECS is a compact object composed of a condensed gas of interacting bosons where the gravity is mostly balanced by the pressure that comes from the interactions. In the context of NS, these bosons are formed by the pairing of neutrons.  The characteristic densities of BECS of paired neutrons ($\sim10^{14}-10^{16}$g/cm$^3$) are close to those of NS in the vicinity of their central regions. Thus,  BECS models are appropriated to describe NS cores but not their outer regions \cite{quintero2019self}. At NS cores, neutrons are expected to pair with parallel spins \cite{116,117,118,119,121,124,125}, giving rise to effective vector bosons whose response to the magnetic field of the star may be relevant \cite{quintero2019self}.

Besides the potential applicability of BECS to the description of some stages of the evolution of NS, it is believed that a considerable amount of stellar dark matter may be some sort of bosonic matter \cite{Jaeckel2010,ringwald2012exploring,Belyaev2019}. Whenever the temperature of this matter is low enough, it will condense, leading to the formation of a condensate dark matter background \cite{Li:2012sg,hui2017ultralight,fukuyama2006relativistic,maleki2019deformed,das2018bose,calzetta2005early}, and even to gravitational bound structures \cite{Li:2012sg,Braaten:2019knj}. Boson stars models have been largely exploited in the study of these hypothetical stars (see, for instance, \cite{Li:2012sg,Braaten:2019knj,Shapiro2012,Davidson2016,Zhang2018,Delgado2020,Ellis2018,Belyaev2019,PhysRevD.105.023001,https://doi.org/10.48550/arxiv.2112.14231}). In this context, the versatility of BECS comes to play since they allow to consider several kinds of bosons (charged/neutral, scalar/vector, relativistic/Newtonian) and situations (zero or finite temperature, magnetic field) starting from the same theoretical setup \cite{latifah2014bosons,quintero2019self}.

The equations of state (EoS) of BECS, as well as their mass and size, are governed by the boson mass $m$ and the interaction strength through the scattering length $a$, although thermal and magnetic effects may be also relevant \cite{latifah2014bosons,Gruber2014, quintero2019self,quintero2019magnetized}. We have studied BECS under the action of a uniform and constant magnetic field \cite{quintero2019self,quintero2019magnetized} and found that, in general, these stars are
spheroidal (because the magnetic field splits the pressure into two components, parallel and perpendicular to its axis), less massive and smaller than the non-magnetic ones. These effects begin to be noticeable for magnetic fields intensities $B\sim10^{-2}B_c$ or higher, being $B_c$ the bosons critical field\footnote{The critical field is defined as the field at which the magnetic energy of the particle equals its mass.}. They are more
relevant at low densities and depend on whether the magnetic field is constant or varies with the density.

On the other hand, thermal effects have been traditionally ignored in the EoS of compact or exotic stars. In the case of fermion stars, the main argument behind the zero temperature approximation is that given their high densities, the Fermi energy is much higher than any temperature achievable inside the star \cite{Camezind}. However, it has been recently show that even the canonical models of neutron stars allow temperatures that range from $10$ to $100$~MeV ($\sim 10^{11}-10^{12}$~K) for densities around $10^{14}-10^{15}$~g/cm$^3$ \cite{Dexheimer2013,Franzon2016,Wey2021}. The effects of these temperatures are nothing but negligible: they change the internal composition and the macroscopic properties of those stars.

Boson systems are much more sensitive to environmental changes (variations in particle density, temperature, and magnetic field) due to the Bose-Einstein condensation \cite{Lismary2018,de2019bose,quintero2021}. In the case of BECS with non-relativistic EoS, a finite temperature increases the pressure at low densities, resulting in an augmentation of the size and mass of the stars \cite{latifah2014bosons,Gruber2014}. When compared to the magnetized BECS described above, it appears, at first glance, that the thermal and magnetic effects oppose each other and that they should cancel out. But, as we shall see, this is not the case.

Our goal in this paper is to study the joint role of temperature and magnetic field on the EoS and the macroscopic properties of BECS. Contrary to previous works on hot BECS \cite{latifah2014bosons,Gruber2014}, we consider the bosons as spin-one relativistic particles. Spin-one bosons are of interest due to their unique magnetic properties in connection with the condensate (see  \cite{yamada1982thermal,Lismary2018,quintero2019self,quintero2021} for a discussion on the so-called Bose-Einstein ferromagnetism and its possible applications). They may arise as bound states of spin-one half fermions, as in the core of NS, but also as fundamental particles in the context of the Standard Model and its extensions \cite{Belyaev2019,delnobile2017direct}.

Regarding magnetized BECS, the consideration of relativistic bosons is required to correctly account for the effects of the super-strong magnetic fields that may be produced in astrophysical environments (up to $10^{18}$~G in the core of NS \cite{Lattimer:2000nx}). For such high intensities, the magnetic field energy might easily become comparable with the rest mass and the thermal energy of the particles and the magnetized vacuum contributes to the system with non-negligible energy, pressure, and magnetization \cite{Quintero2017PRC,quintero2021}. 

The paper is organized as follows: We devote Section \ref{sec2} to obtaining and studying the EoS of magnetized BECS at finite temperatures. In Section \ref{sec3} we present the structure equations that we use to obtain the macroscopic properties of these stars. Later on in this section, we discuss the dependence on temperature of the properties of non-magnetized BECS, with emphasis on the stars' stability and evolution, and finally, we analyze the interplay of the thermal and magnetic effects. Concluding remarks are given in Section \ref{sec5}.

We use natural units $\hbar=c=1$, and for numerical calculations we consider a spin-one boson with mass $m=2m_N$ and magnetic moment $\kappa=2\mu_N$, being $m_N$ and $\mu_N$ the mass and the magnetic moment of the neutron. For such masses, and scattering lengths ranging from $a=1$ to $15$~fm, the model at zero temperature and magnetic field gives stars with maximum masses around $0.6-2.5 M_{\odot}$, where $M_{\odot}$ is the mass of the Sun, and maximum central densities around $\rho_c\sim10^{16}-10^{15}$g/cm$^3$ respectively \cite{chavanis2012bose,latifah2014bosons}. In the plots, we use $a=13$~fm.  The choice of these values is only for practical purposes, being the discussions of our results valid for any hot BECS with a magnetic field.

\section{Equations of state of magnetized  Bose-Einstein condensate stars at finite temperature}
\label{sec2}
The Hamiltonian of a gas of interacting bosons can be written as a sum of an ideal gas Hamiltonian $\hat{H}_{th}$ that includes the effects of temperature and any external field, plus the particle-particle interaction Hamiltonian $\hat{H}_{int}$,
\begin{equation}\label{hamiltonian}
	\hat{H} = \hat{H}_{int} + \hat{H}_{th}.
\end{equation}
Then, the grand thermodynamic potential per unit volume of the system is \cite{latifah2014bosons,quintero2019self}
\begin{equation}\label{omega1}
	\Omega = \Omega_{int} + \Omega_{th}.
\end{equation}

The boson-boson interaction is considered as a two-body contact interaction, i.e., we assume that only the low energy binary collisions are relevant \cite{chavanis2012bose,latifah2014bosons,quintero2019self,Li:2012sg}.
This approximation holds as far as the quantum fluctuations are negligible (mean field approximation), but this is to expect for a dense system as the BECS ($\rho\sim10^{14}-10^{15}$g/cm$^3$) at the relatively low temperatures needed for the existence of the condensate.

Despite its simplicity, the assumption of a two-body contact interaction allows to obtain thermodynamic consistent EoS for a gas of interacting bosons at finite temperature \cite{latifah2014bosons} or magnetic field \cite{quintero2019self}, whose corresponding mass-radius curves have maximum values and shapes that are consistent with other BECS models \cite{chavanis2012bose,Gruber2014}. Hence, this is a useful starting point to understand the physics of hot magnetized BECS.

Under this assumption, $\Omega_{int}$ transforms into \cite{latifah2014bosons}
\begin{equation}\label{omega0}
	\Omega_{int}=1/2u_0 \rho^2_T,
\end{equation}
where $u_0 = 4 \pi a/m$ indicates the strength of the interaction and $\rho_T$ is the density of particles and antiparticles, although as we shall see, the effects of antiparticles are negligible.

We compute $\Omega_{th}$ starting by its definition in the one-loop approximation \cite{quintero2021}
\begin{eqnarray}\label{omega2}
	\Omega_{th}(\mu,T,B)&=& \Omega_{st}(\mu,T,B)+\Omega_{vac}(B)\nonumber\\&=& \sum_{s} \int\limits_{0}^{\infty}\frac{p_{\perp}dp_{\perp}dp_3}{(2\pi)^2 \beta} \ln \left( f_{BE}^{+} f_{BE}^{-}\right) + \sum_{s}\int\limits_{0}^{\infty}\frac{p_{\perp}dp_{\perp}dp_3}{(2\pi)^2}\varepsilon(p_3,p_{\perp} B,s).
\end{eqnarray}
Here $\mu$ is the chemical potential, $\beta=1/T$ is the inverse of the temperature $T$, $B$ is the magnetic field intensity and $f_{BE}^{\pm} = \left [1-e^{-(\varepsilon\mp \mu)\beta} \right ] $ stands for particles/antiparticles. The sum over $s=0,\pm 1$ contains the spin states, $p_3$ and $p_{\perp}$ are the momentum components parallel and perpendicular to the magnetic field axis, and $\varepsilon = \sqrt{m^2+p_3^2+p_{\perp}^2-2\kappa s B\sqrt{p_{\perp}^2+m^2}}$ is the energy spectrum of the vector bosons under an external magnetic field in the $z$ direction \cite{Quintero2017PRC}. We would like to remark that, microscopically, the magnetic field can be always considered as locally uniform and constant.

The first integral in Eq.~(\ref{omega2}),  $\Omega_{st}$, accounts for the statistical contribution of particles and antiparticles, and depends on $T$, $B$ and $\mu$. Using the Taylor expansion of the logarithm and integrating over the momentum components $\Omega_{st}$ can be rewritten as \cite{quintero2021}
\begin{eqnarray}\label{omega4}
	\hspace{-0.5cm}\Omega_{st}(b,\mu,T)= -  \sum_{s} \sum_{n=1}^{\infty} \frac{e^{n \mu \beta}+e^{- n \mu \beta }}{ 2 \pi^2 n \beta } \left \{ \frac{y_0^2}{n\beta^2} K_2 (n \beta y_0)
	- \alpha \int\limits_{y_0}^{\infty} dx
	\frac{x^2}{\sqrt{x^2+\alpha^2}} K_1 (n \beta x) \right \},
\end{eqnarray}
\noindent where $K_l(x)$ is the McDonald function of order $l$, $y_0= m \sqrt{1-s b}$, $\alpha=s m b/2$, and $b=\frac{B}{B_c}$ with $B_c=\frac{m}{2\kappa}$ the critical magnetic field of neutral vector bosons (for the numerical values of $m$ and $\kappa$ that we will consider $B_c=7.8 \times 10^{19}$~G). Note that Eq.~(\ref{omega4}) holds for any temperature \cite{quintero2021}.

The second integral in Eq.~(\ref{omega2}) is the zero-point energy or vacuum term. It is independent of $T$ and $\mu$ and has an ultraviolet divergence. After renormalization  it  reads \cite{Quintero2017PRC,quintero2021}
\begin{align}\label{omegavac}
	\Omega_{vac}(b) = -\frac{m^4}{288 \pi}\left( b^2(66-5 b^2)
	-3(6-2b-b^2)(1-b)^2 \log(1-b)\right.
	\\\nonumber
	\left. -3(6+2b-b^2)(1+b)^2\log(1+b) \right).
\end{align}

The thermodynamic potential of the interacting boson gas at finite temperature is obtained by adding Eqs.~(\ref{omega0}), (\ref{omegavac}) and (\ref{omega4})  as $\Omega=\Omega_{int} + \Omega_{st}+\Omega_{vac}$. Thus, the EoS  of the BECS can be computed as \cite{latifah2014bosons,quintero2019self}
\begin{subequations}\label{EoSdefinition}
	\begin{align}
		P_{\parallel}&= -\Omega + \rho_T \left(\frac{\partial \Omega}{\partial \rho_T}\right)_{\mu,T,B}-\frac{B^2}{8 \pi}, \\
		P_{\perp}&= -\Omega + \rho_T \left(\frac{\partial \Omega}{\partial \rho_T}\right)_{\mu,T,B} - M B +\frac{B^2}{8 \pi}, \\	
		E &= \Omega + \mu \left(\frac{\partial \Omega}{\partial \mu}\right)_{T,B}  - T \left(\frac{\partial \Omega}{\partial T}\right)_{\mu,B}+\frac{B^2}{8 \pi}, \\
		\rho &= \rho_{gs} - \left(\frac{\partial \Omega}{\partial \mu}\right)_{T,B},\\
		\mathcal{M}&=-\left(\frac{\partial\Omega}{\partial B}\right)_{T,\mu},
	\end{align}
\end{subequations}
where $P_{\parallel}$ and $P_{\perp}$ are the pressures along and perpendicular to the magnetic axis, $E$ is the internal energy, $\mathcal{M}$ is the magnetization, and $\rho$ is the particle density of the system. Note $\rho$ includes the condensed particles $\rho_{gs}$ as well as the difference of the particles/antiparticles densities in the excited states $- \left(\frac{\partial \Omega}{\partial \mu}\right)_{T,\mu}$.

After derivation and simplification, the EoS of the hot magnetized BECS read
\begin{subequations}\label{EoSR}
	\begin{align}
		P_{\parallel}&=\frac{1}{2}u_0\rho_T^2 - \Omega_{st} - \Omega_{vac} -\frac{B^2}{8\pi} ,\label{EoSPpar}\\
		P_{\perp}&=\frac{1}{2}u_0\rho_T^2 - \Omega_{st} - \Omega_{vac} - 	{\mathcal M}^{R} B +\frac{B^2}{8\pi} , \label{EoSPper} \\
		E&=\frac{1}{2}u_0\rho_T^2 + m\rho_{gs} + E_{st} + \Omega_{vac} + \frac{B^2}{8\pi} \\
		\rho&=\rho_{gs} +\rho^{+}-\rho^{-},\\
		{\mathcal M}&=\frac{\kappa}{\sqrt{1-b}}\rho_{gs}+{\mathcal M}_{st}+{\mathcal M}_{vac},
	\end{align}
\end{subequations}
with
\begin{equation}
	E_{st} = -\Omega_{st}+\sum_{s} \sum_{n=1}^{\infty}  \frac{e^{n \mu \beta}+e^{- n \mu \beta }}{n}  \left \{ \frac{ y_0^3 T }{4 \pi^2}[ K_1 (n \beta y_0)+K_3 (n \beta y_0)] \right.+ \frac{\alpha n}{2 \pi^2}\left. \int\limits_{y_0}^{\infty} dx
	\frac{x^3}{ \sqrt{x^2+\alpha^2}} K_0 (n \beta x) \right \},
\end{equation}

\begin{equation}
	\rho^{\pm}=\sum_{s} \sum_{n=1}^{\infty} \frac{e^{\pm n \mu \beta}}{2 \pi^2} \left \{ \frac{y_0^2 }{n\beta} K_2 (n \beta y_0)  + \alpha \int\limits_{y_0}^{\infty} dx
	\frac{x^2}{\sqrt{x^2+\alpha^2}} K_1 (n \beta x) \right \},
\end{equation}
and
\begin{eqnarray}\label{magst}
	{\mathcal M}_{st} = \sum_{s} \frac{\kappa s}{\pi^2 \beta } \sum_{n=1}^{\infty}  \frac{e^{n \mu \beta}+e^{- n \mu \beta }}{n} \left \{ \frac{m y_0 }{ (2-b s)} K_1 (n \beta y_0) +\int\limits_{y_0}^{\infty} dx
	\frac{x^4}{ 2 (x^2+\alpha^2)^{3/2}} K_1 (n \beta x) \right \},  \quad \quad
\end{eqnarray}
\begin{eqnarray}\label{magvac}
	{\mathcal M}_{vac}=-\frac{\kappa m^3}{72 \pi} \left \{ 7 b (b^2-6) -3(2b^3-9b+7)\log(1-b) -3(2b^3-9b-7)\log(1+b)\right \}. \quad \quad
\end{eqnarray}
Here $\rho^{+}/\rho^{-}$ are the particle/antiparticle density respectively.

\begin{figure}[h!]
	\centering
	\includegraphics[width=0.49\linewidth]{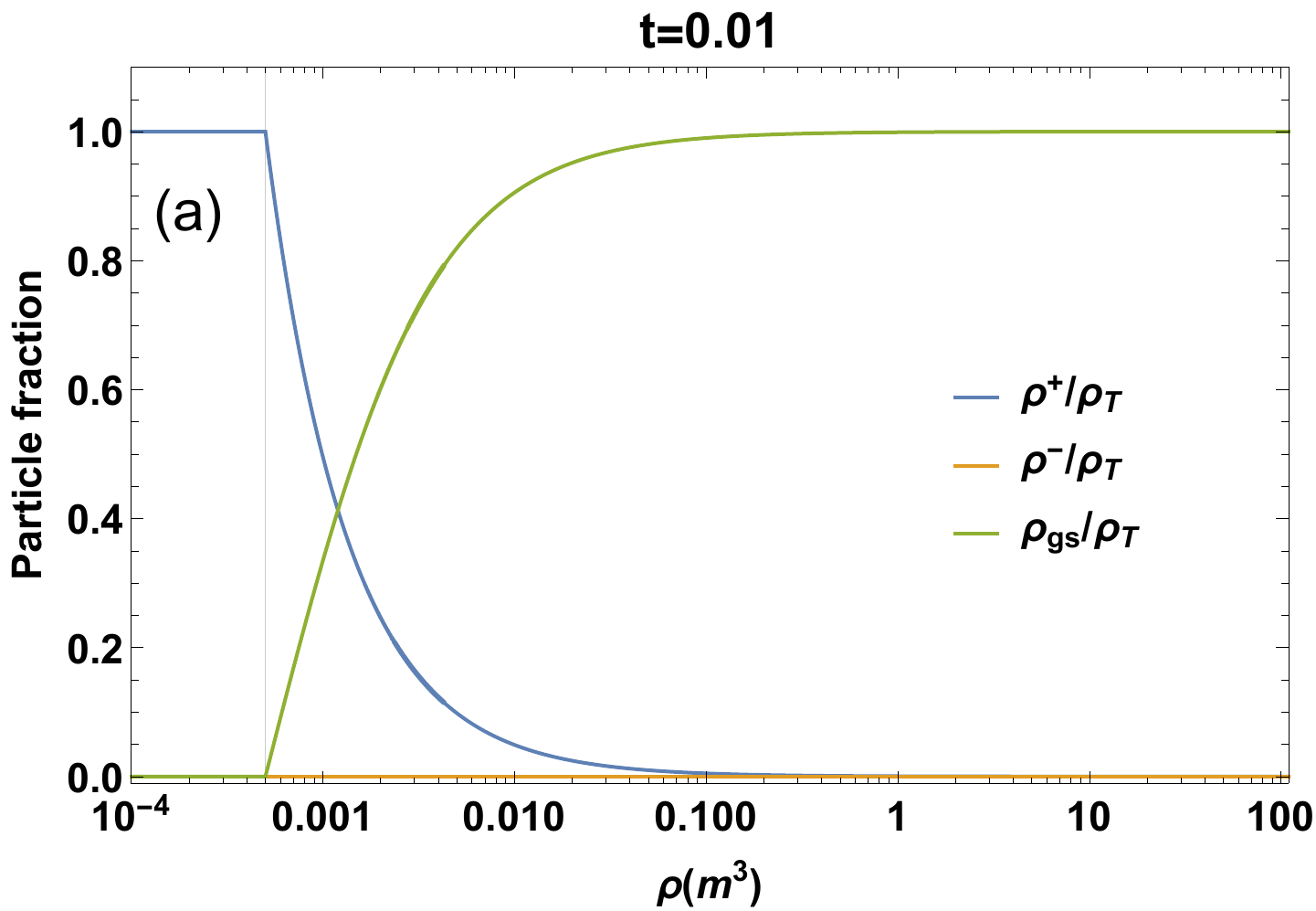}
	\includegraphics[width=0.49\linewidth]{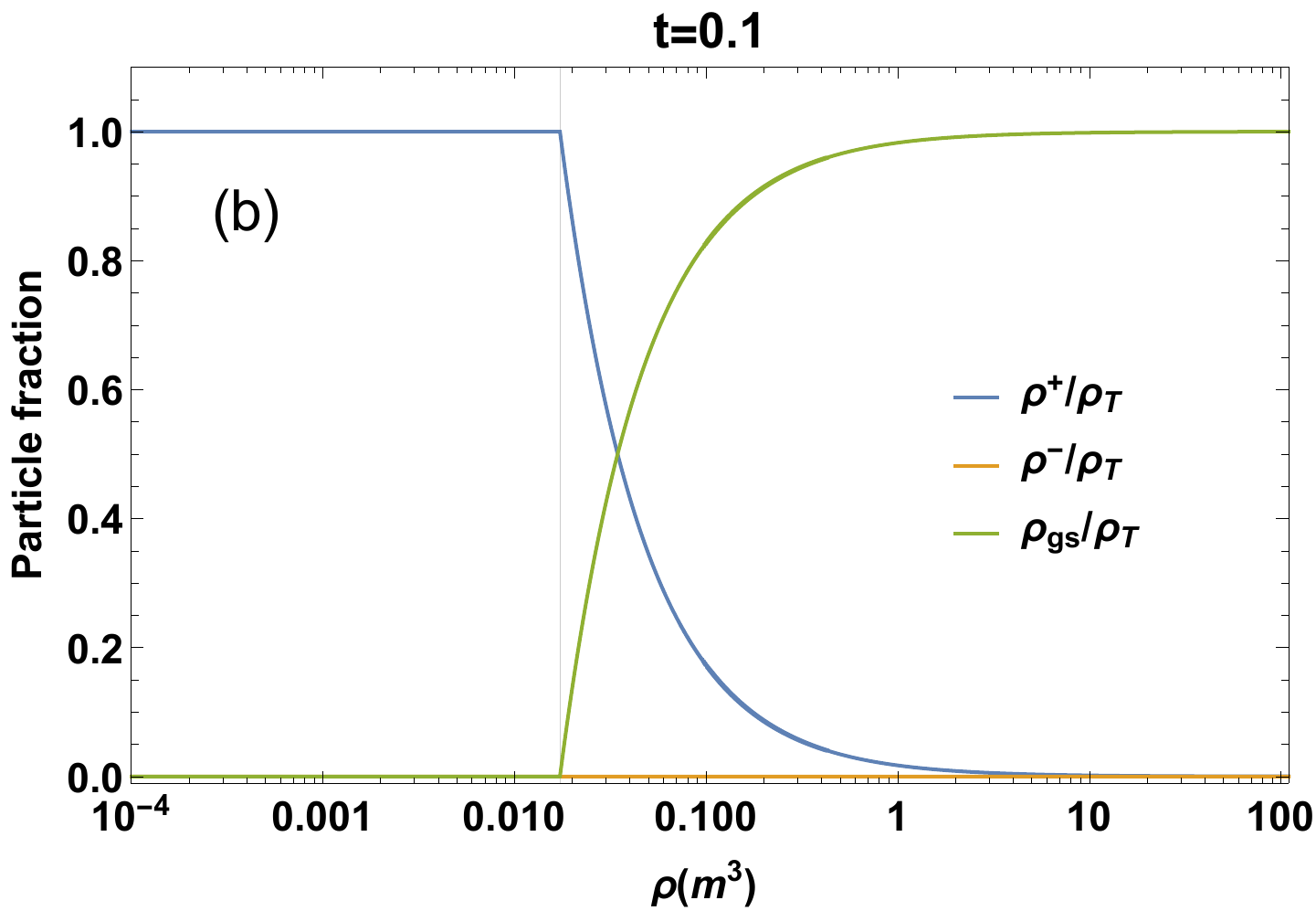}
	\includegraphics[width=0.49\linewidth]{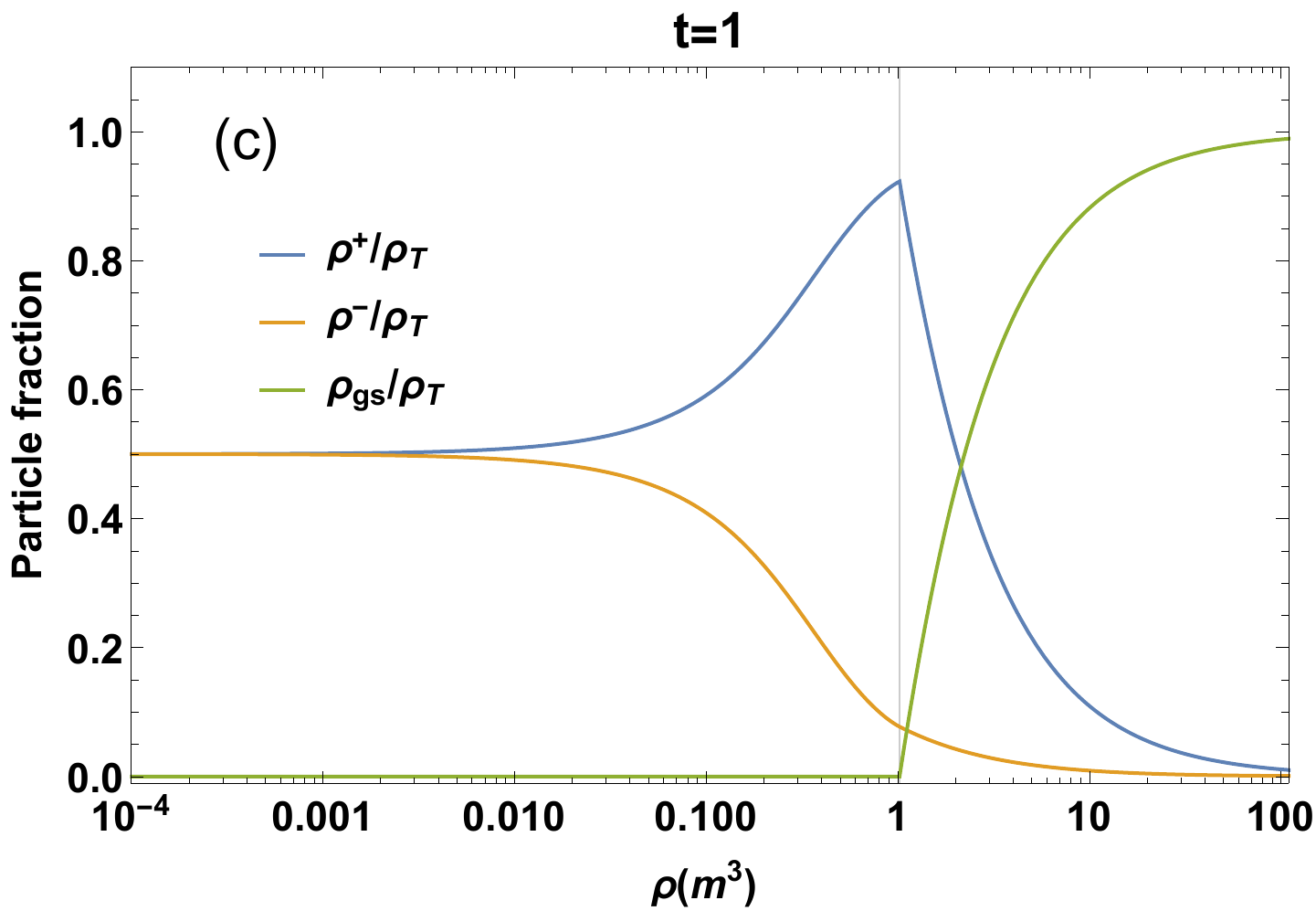}
	\includegraphics[width=0.49\linewidth]{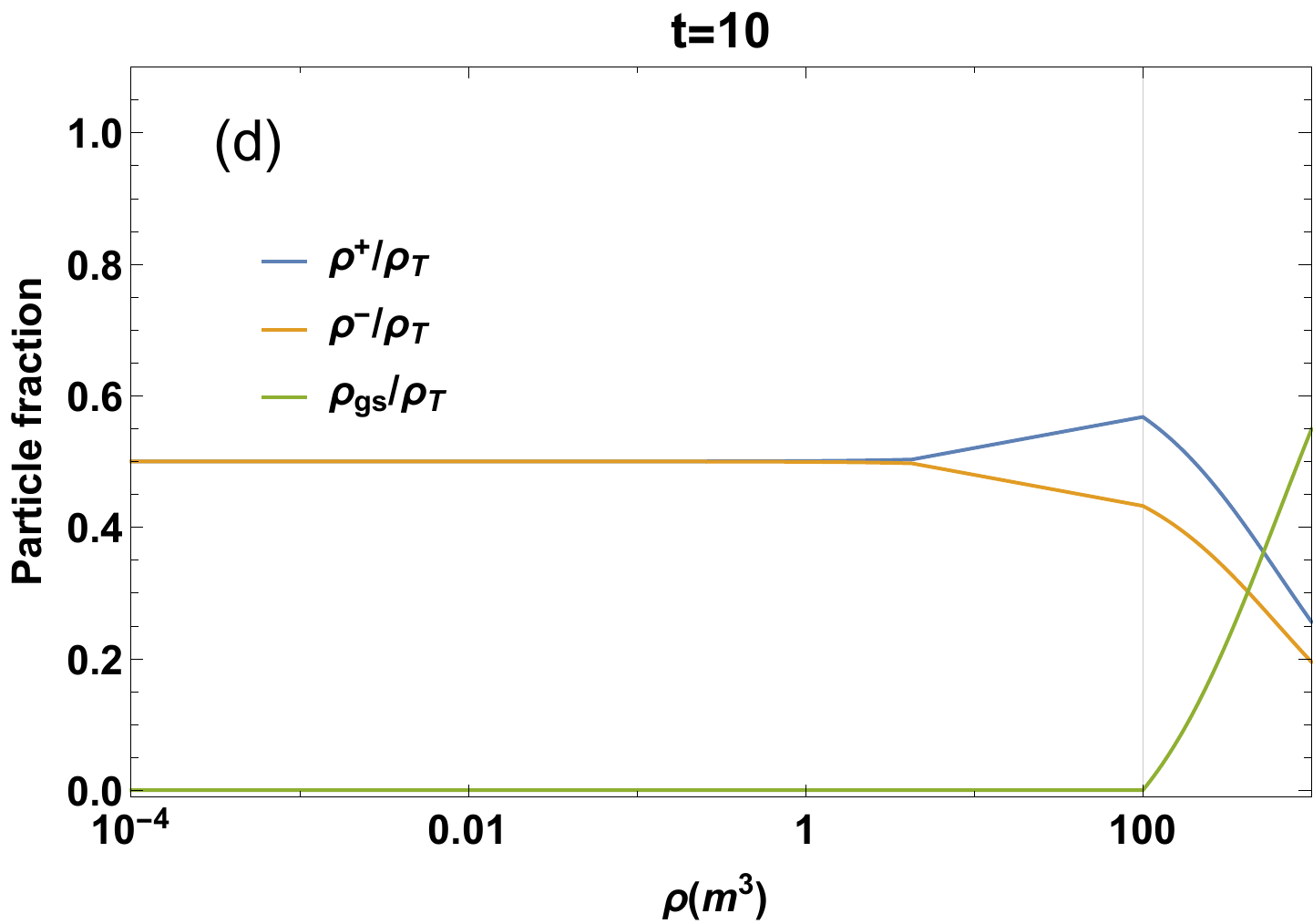}	
	\caption{\label{F1} Fraction of particles, antiparticles, and particles in the ground state as functions of $\rho$ for several values of the temperature. The vertical lines correspond to the density at which the condensate appears.}
\end{figure}

Let us focus first on the dependence on $T$ of the fraction of particles, antiparticles and condensed particles ($\rho^{+}/\rho_T$, $\rho^{-}/\rho_T$ and $\rho_{gs}/\rho_T$ respectively) to determine the range of temperatures to use in our study of BECS. In Fig.~\ref{F1} we show those quantities as functions of $\rho$ for several values of the temperature in units of mass $t=T/m$ and $B=0$. The vertical lines mark the critical density of condensation: for densities at their right, a fraction of the gas is in the condensed state. The region where the Bose-Einstein condensation exists decreases with the increase in temperature.
For $m=2m_N$, $\rho=m^3$ is equivalent to $\sim 3 \times 10^{18}$g/cm$^3$ which is two orders above the typical central densities of these BECS ($10^{15}-10^{16}$g/cm$^3$, i.e. $\sim 0.001-0.01 m^3$). Hence, for $T \geq 0.1m$ ($\sim 10^{11}$~K) the gas does not condense at any density that could exist inside a stable star. In this regard, we will restrict our analysis of hot magnetized BECS to temperatures such that $T\leq0.01 m$, to guarantee the existence of Bose-Einstein condensation inside the star.

In the selected range of temperatures, the antiparticle density is negligible and they will not have any relevant influence on the hot BECS physics.
This situation does not change at a finite magnetic field, since to affect antiparticle production and Bose-Einstein condensation, $B$ needs to be above $\sim0.1 B_c$, which for $B_c=7.8 \times 10^{19}$~G is an order above the highest values of the magnetic field expected in compact stars \cite{quintero2021}.

\begin{figure}[h!]
	\centering
	\includegraphics[width=0.49\linewidth]{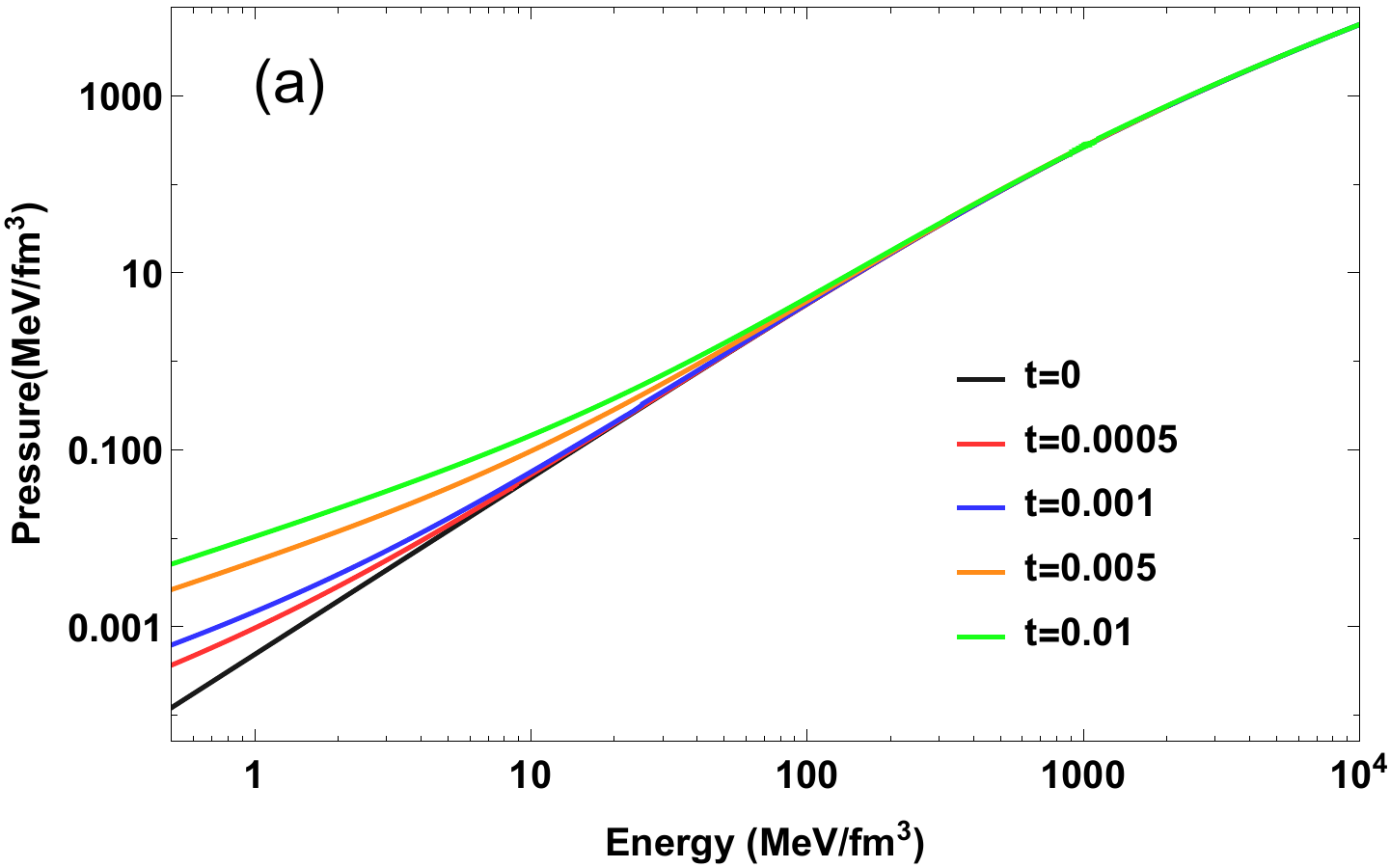}
	\includegraphics[width=0.49\linewidth]{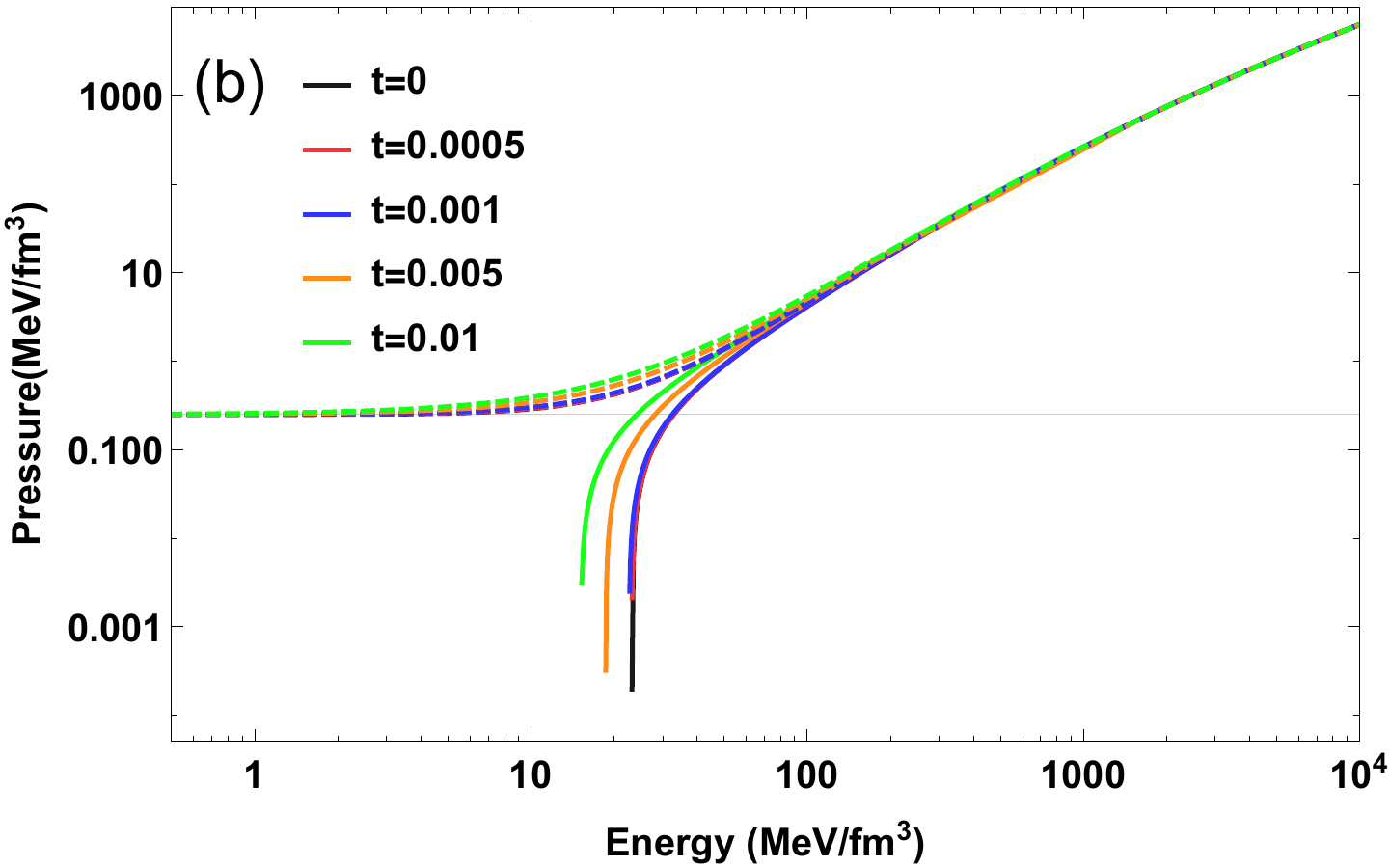}
	\caption{\label{F2} The EoS of hot BECS without (a) and with magnetic field $B=10^{17}$~G (b). In the right plot, the dashed and solid lines correspond to the parallel and perpendicular pressures respectively.}
\end{figure}

With the antiparticles out of the picture, at $B=0$ the main effect of a finite temperature in the BECS EoS is an increase of the pressure for the lower densities ($E\leq100$MeV/fm$^3$, $\rho\leq10^{14}$g/cm$^3$). This happens because as $T$ increases from $0$, the thermal pressure $-\Omega_{st}(b,\mu, T)$ starts to gain relevance in comparison to the other components of the pressure. At high densities the boson-boson interaction pressure ($ u_0\rho_T^2/2$ in Eq.~(\ref{EoSPpar})-(\ref{EoSPper})) prevails, being the total pressure corresponding to those densities barely affected by the temperature (Fig.~\ref{F2}a). 

Turning on the magnetic field makes the pressure anisotropy noticeable (Fig.~\ref{F2}b). In this case, as the density decreases, the parallel pressure Eq.~(\ref{EoSPpar}) (dashed lines) tends to $B^2/8\pi$ and becomes density-independent, while the perpendicular pressure Eq.~(\ref{EoSPper}) (solid lines) goes to $-B^2/8\pi$ becoming negative. This instability imposes a lower bound on the densities that can exist inside the star for a given magnetic field \cite{quintero2019self}. 

The anisotropy in the pressure and the instability caused by the magnetic field persist for all the considered temperatures. The magnetic energy per boson in units of mass at $B=10^{17}$~G, $~\kappa B \sim 10^{-4}m$, is lower than all the used temperatures $5\times10^{-4}$--$10^{-2}m$, except when $T=0$. This may suggest that the increase in thermal pressure should be enough to balance the instability. However, what sets the anisotropy on is the term corresponding to the Maxwell classical energy of the magnetic field: $B^2/8\pi$. This term adds, or subtracts, to the pressures an energy per boson $B^2/8\pi\rho$ that is almost negligible at high densities, $B^2/8\pi\rho\sim10^{-5} m$ for $\rho\sim 10^{16}$~g/cm$^3$ and $E\sim 10^4$~MeV/fm$^3$, but that leads the system behavior at the low ones: $B^2/8\pi\rho\sim10^{-2}m$ for $\rho\sim10^{13}$~g/cm$^3$ and $E\sim 10$~MeV/fm$^3$. Hence, the Eos in Fig.~\ref{F2}b are dominated by the magnetic field at low densities, by the temperature at the intermediate ones, and by the interactions at the highest values of $\rho$. So, the selected values of $T$ and $B$ guarantee that we will study stars where their effects are both relevant, although dependent on the density.

\section{Macroscopic properties of relativistic Bose-Einstein condensate stars at finite temperature under the action of a magnetic field}
\label{sec3}

Now we focus on both, thermal and magnetic effects, on the macroscopic properties of BECS. To do so, we will assume that the temperature is constant inside the star. The temperature of compact stars is thought to increase toward the center \cite{Wey2021,Franzon2016}. However, detailed calculations of the inner profiles of NS with realistic NS EoS show that the increase of $T$ is usually no longer than one order and that the biggest change takes place near the surface of the star \cite{81,251}, thus considering a constant temperature is a good approximation. Similarly, self-consistent numerical calculations with realistic models for the inner magnetic fields of white dwarfs and neutron stars show that variations in the magnetic field intensity inside these objects do not usually exceed one order of magnitude \cite{71,247,251}. Hence, we also assume that, inside the star, the magnetic field is uniform and constant.

As we showed in the previous section, the magnetic field splits the inner pressure of the star into two components, one along and the other perpendicular to the magnetic axis. Since the inner pressure of compact objects is, in general, proportional to its radius \cite{Lattimer:2000nx}, magnetized stars are not spherical but axially deformed.  As a consequence, the macroscopic structure of a magnetized compact object can not be accounted for with the standard Tolman-Oppenheimer-Volkoff (TOV) equations \cite{Camezind} because they describe static spherically symmetric stars and do not admit a pressure anisotropy of the type caused by the magnetic field. To properly take into account the anisotropy, we will use the so-called $\gamma$-structure equations given as \cite{Terrero:2018utx}
\begin{subequations}\label{gTOV}
	\begin{eqnarray}
		&& \frac{dM}{dr}= \gamma r^{2}\frac{(E_{\parallel} +E_{\perp})}{2}, \label{gTOV1}\\
		&&\frac{dP_{\parallel}}{dr}=-\frac{(E_{\parallel}+P_{\parallel})[\frac{r}{2}+4 \pi G r^{3}P_{\parallel}-\frac{r}{2}(1-\frac{2GM}{r})^{\gamma}]}{ r^{2}(1-\frac{2GM}{r})^{\gamma}}, \label{gTOV3}\\
		&&\frac{dP_{\perp}}{dr}=-\frac{(E_{\perp}+P_{\perp})[\frac{r}{2}+4\pi Gr^{3}P_{\perp}-\frac{r}{2}(1-\frac{2GM}{r})^{\gamma}]}{ r^{2}(1-\frac{2GM}{r})^{\gamma}}. \label{gTOV2}
	\end{eqnarray}
\end{subequations}
These equations establish the hydrodynamic equilibrium between the gravity and the internal pressure of a spheroidal compact object, provided it is close to the spherical shape ($\gamma \simeq 1$) \cite{Terrero:2018utx}.

In Eqs.~(\ref{gTOV}) $G$ is Newton's gravitational constant and $M(r)$ is the mass enclosed in the spheroid of equatorial radius $r$. To obtain the total mass and radii of the star, Eqs.~(\ref{gTOV}) are integrated with the initial conditions $E_0 = E(r=0)$, $P_{\parallel_0} = P_\parallel(r=0)$, and $P_{\perp_0} = P_\perp(r=0)$, where $E_0$, and $P_{\perp_0}$ and $P_{\parallel_0}$ are taken from the EoS, while the condition $P(R) = 0$ defines the star equatorial radius from which the total mass $M(R)$ is computed. At each integration step, $E_{\parallel}$ and $E_{\perp}$ are computed using the parametric dependence of the energy in each pressure derived from  Eqs.~(\ref{EoSPpar})-(\ref{EoSPper}).

The parameter $\gamma$ accounts for the axial deformation of the star and relates the polar and equatorial radius as $z=\gamma r$ ($Z=\gamma R$) such that if $\gamma =1$ the star is spherical while for $\gamma>1$ ($\gamma<1$) it is a prolate (oblate) spheroid. In Eqs.~(\ref{gTOV}) $\gamma$ acts as an external parameter \cite{Terrero:2018utx}. To solve them, in \cite{Terrero:2018utx} $\gamma$ has been considered as the ratio between the parallel and the perpendicular central pressures,
\begin{equation}
	\gamma=P_{\parallel_0}/P_{\perp_0}. \label{gamma}
\end{equation}
This \textit{anzats} connects the system's geometry with its physical properties and follows from the proportionality between the radius and the central pressure of spherical stars \cite{Lattimer:2000nx}. When $B=0$, $P_{\perp}=P_{\parallel}$, $\gamma=1$, Eqs.~(\ref{gTOV}) reduced to TOV equations and the spherical case is recovered. Regardless their approximated character, the study of the macroscopic structure of magnetized compact objects through the combination of Eqs.~(\ref{gTOV}) with the \textit{ansatz} Eq.~(\ref{gamma}) yields reasonable results \cite{Terrero:2018utx,quintero2019self,quintero2019magnetized,momentofinertia,Samantha}, that are qualitatively similar to those coming from models that consider more sophisticated magnetic field geometries \cite{Rizaldy2018,Chatterjee,81,251,247,71}.

Note that Eqs.~(\ref{gTOV}) guarantee the hydrodynamic equilibrium for spheroidal stars, but equilibrium does not necessarily imply stability  \cite{Camezind,shapiro2008black,gleiser1988stability,Samantha}. We use two criteria to study the stability of the solutions. The first one guarantees stability against radial oscillations and requires that $dM/d\rho\geq0$ \cite{Camezind}. The second criterion request $M(\rho)<M_B(\rho)$, where $M_B$ is the baryonic mass of the star Eq.~(\ref{MB}). The last assures that the stars are stable against the dispersion of their particles \cite{Camezind,gleiser1988stability}.
\begin{eqnarray}\label{MB}
	M_{B}&=&m\int_0^R \frac{4\pi r^2 \rho(P(r)) }{(1-\frac{2GM(r)}{r})^{\gamma/2}}dr.
\end{eqnarray}

Since the solution of Eqs.~(\ref{gTOV}) gives two radii, to compare with the $B=0$ case it is convenient to use the mean radius $R_m$ so that the surface of the sphere it determines is equal to the surface of the spheroidal star
\begin{equation}
	A=2 \pi R \left (R+\frac{Z}{\epsilon} \arcsin \epsilon \right ),
\end{equation}
where $\epsilon=\sqrt{1-\gamma}$ is the ellipticity \cite{Samantha}.

Apart from the mass and radii, we are also interested in analyzing the thermal and magnetic effects over the compactness $G M/R$, the gravitational redshift \cite{Samantha}
\begin{eqnarray}
	z&=&\frac{1}{(1-\frac{2GM}{R})^{\gamma/2}}-1,
\end{eqnarray}
and the mass quadrupolar moment \cite{Samantha}
\begin{equation}\label{Q}
	Q=\frac{\gamma}{3}(M/M_{\odot})^3(1-\gamma^2),
\end{equation}
for the case of hot magnetized stars, since for spherical stars $Q=0$. In the case of spherical non-magnetized stars at finite temperature we show the moment of inertia $I$ instead
\begin{eqnarray}\label{I1}
	I=\int_0^R 4\pi r^4 E(P(r))dr.
\end{eqnarray}

\subsection{Non-magnetized Bose-Einstein condensate stars at finite temperature}

\begin{figure}[h!]
	\centering
	\includegraphics[width=0.49\linewidth]{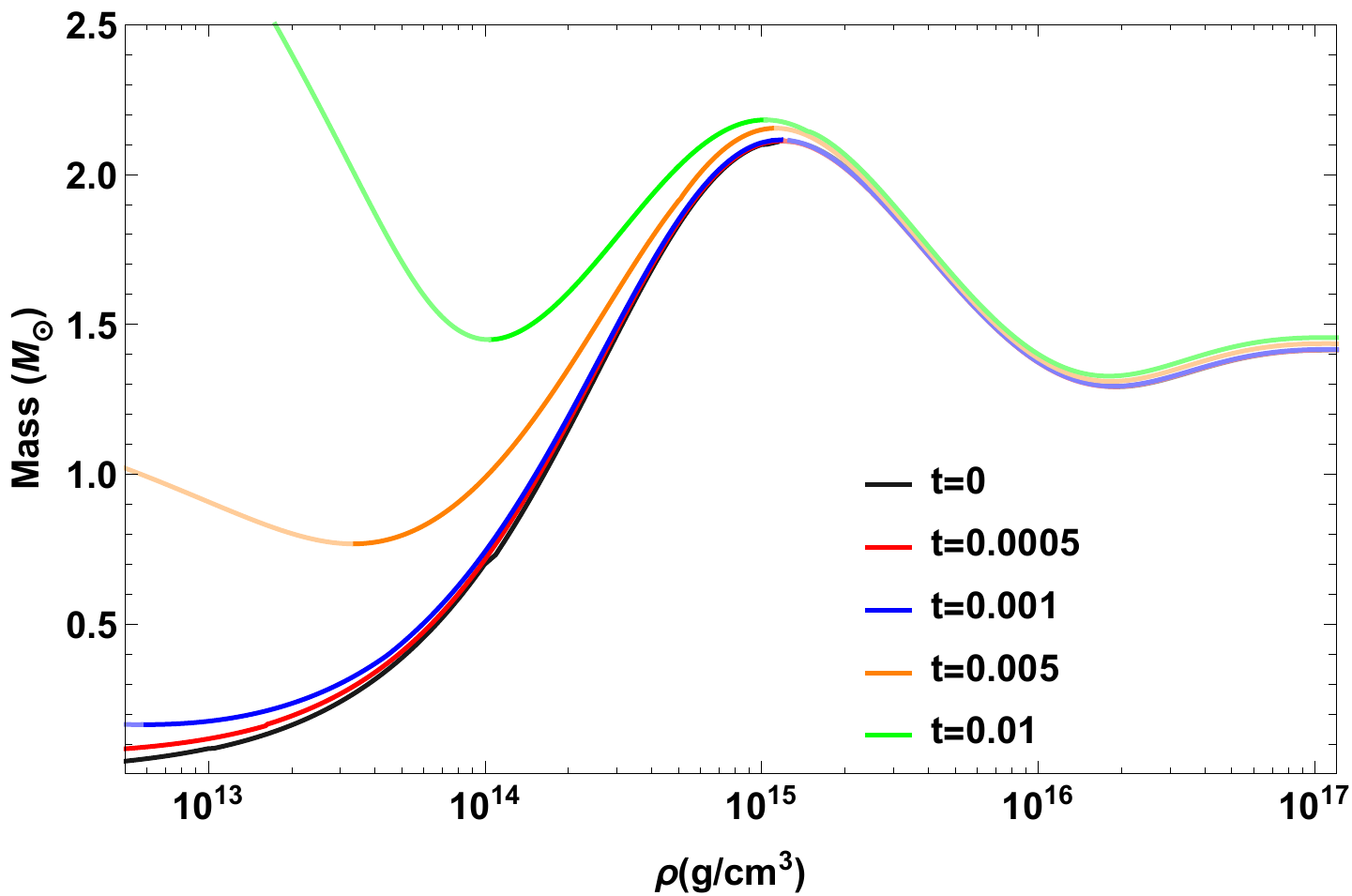}
	\caption{\label{F3} The mass vs. the central mass density of non-magnetized BECS with different inner temperatures.}
\end{figure}

We will first concentrate on understanding the effects of temperature on non-magnetized BEC stars. Fig.~\ref{F3} shows the masses of the stars that result from solving the $\gamma$-structure equations with the EoS at $B=0$ (Fig.~\ref{F2}a). The increase that the temperature causes in the pressure at low densities provokes an increase in the masses of the stars compared to the zero temperature case. This behavior have been observed for non-relativistic BECS and white dwarfs in \cite{latifah2014bosons} and \cite{peterson2021effects} respectively.  What we found unexpected is that the temperature affects the stability of BECS.

In Fig.~\ref{F3} the unstable sections of the curves have a lighter color. The stability analysis reduced to looking for the region where $\partial M/\partial \rho\geq0$, since the baryonic mass criterion is always fulfilled for a broader range of central densities. The interval of central densities of stable stars gets smaller as the temperature increases. Table \textbf{I} collects the extreme values of the central mass density and the mass of the stable stellar configurations for each temperature. The variation with $T$ of the maximum mass and its corresponding central density is almost negligible, while the minimum mass and central density remarkably increase with the temperature. Hence, one expects that, even if we allow $T>0.01 m$, there would be a temperature beyond which there are no stable stars in the range of central densities considered. At such high temperatures, the statistical pressure $-\Omega_{st}$ becomes so high that a gravitationally bound structure can not form. In the remaining figures of this section, we show only the stable part of the curves.
\begin{table}[h!]\label{table}
	\begin{tabular}{cccccc}
		\cline{2-5}
		\multicolumn{1}{c|}{\multirow{2}{*}{\textbf{}}} & \multicolumn{1}{c|}{$\rho_{min} (10^{11} g/cm^{3})$}                                              & \multicolumn{1}{c|}{$\rho_{max}(10^{15} g/cm^3$)}                                                & \multicolumn{1}{c|}{$M_{min}(M_{\odot})$}                                  & \multicolumn{1}{c|}{$M_{max}(M_{\odot}$)}                                  \\ \cline{1-5}
		\multicolumn{1}{|c|}{\textbf{t=0}}              & \multicolumn{1}{c|}{$-$}                                                          & \multicolumn{1}{c|}{$1.23$}                                                   & \multicolumn{1}{c|}{-}                                             & \multicolumn{1}{c|}{$2.11$}                                      \\ \hline
		\multicolumn{1}{|c|}{\textbf{t=0.0005}}         & \multicolumn{1}{c|}{$3.10$} &
		\multicolumn{1}{c|}{$1.19$} &
		\multicolumn{1}{c|}{$0.05$}   &
		\multicolumn{1}{c|}{$2.11$}
		\\ \hline
		\multicolumn{1}{|c|}{\textbf{t=0.001}}         & \multicolumn{1}{c|}{$9.72$} &
		\multicolumn{1}{c|}{$1.18$}&
		\multicolumn{1}{c|}{$0.16$}   &
		\multicolumn{1}{c|}{$2.11$}
		\\ \hline
		\multicolumn{1}{|c|}{\textbf{t=0.005}}         & \multicolumn{1}{c|}{$336$} &
		\multicolumn{1}{c|}{$1.12$}  &
		\multicolumn{1}{c|}{$0.77$}   &
		\multicolumn{1}{c|}{$2.15$}
		\\ \hline
		\multicolumn{1}{|c|}{\textbf{t=0.01}}          & \multicolumn{1}{c|}{$1034$}		&
		\multicolumn{1}{c|}{$1.03$}      &
		\multicolumn{1}{c|}{$1.45$}   &
		\multicolumn{1}{c|}{$2.18$}
		\\ \hline
		\multicolumn{1}{l}{}                            & \multicolumn{1}{l}{}                       & \multicolumn{1}{l}{}                       & \multicolumn{1}{l}{}                       & \multicolumn{1}{l}{}
	\end{tabular}
	\caption{Extreme values of the central mass density and the total mass of the stable stellar configurations at finite temperature and zero magnetic field.}
\end{table}

\begin{figure}[h!]
	\centering
	\includegraphics[width=0.49\linewidth]{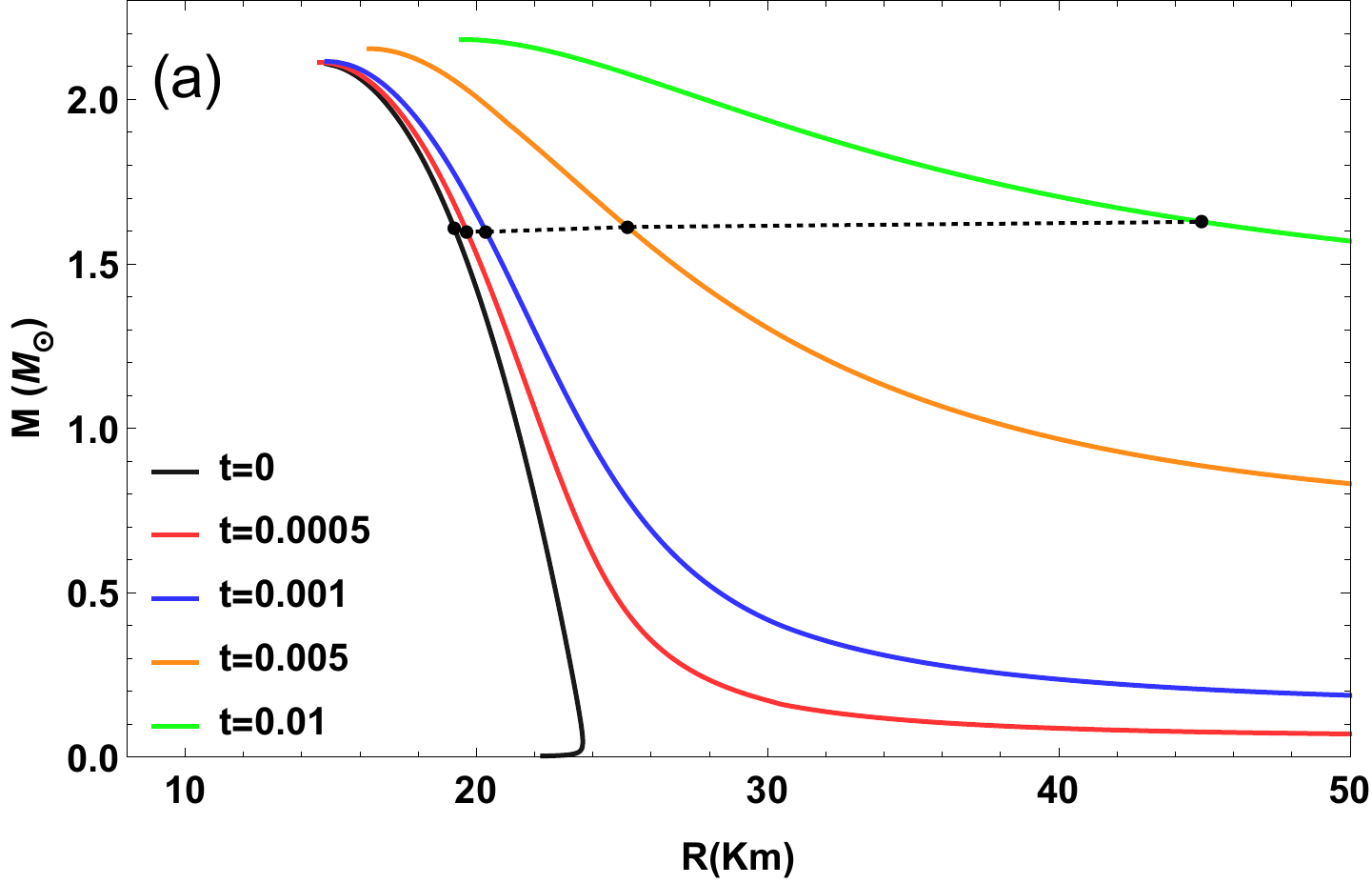}
	\includegraphics[width=0.49\linewidth]{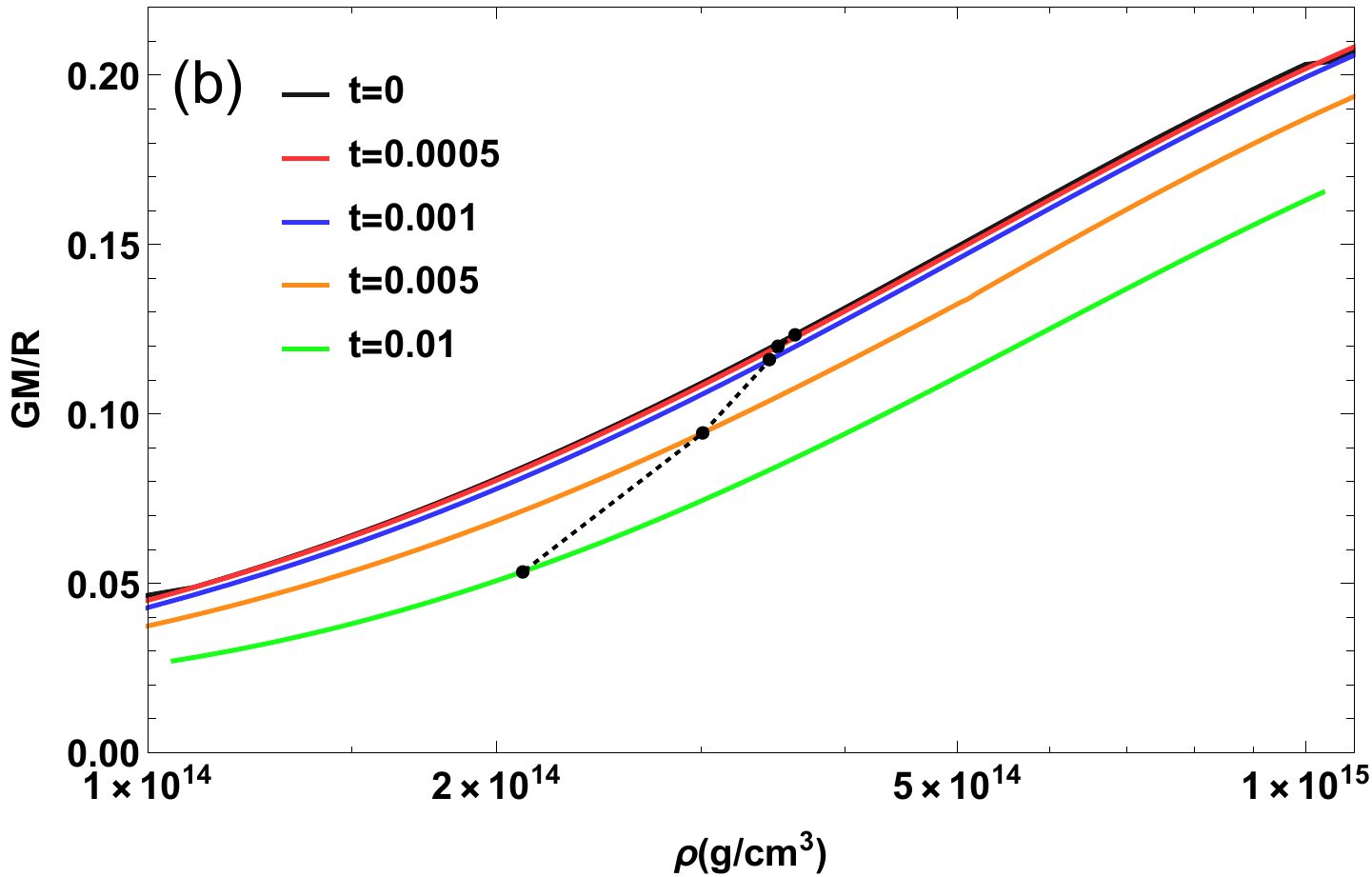}
	\caption{\label{F4} Mass-radius diagram (a) and compactness (b) of non-magnetized BECS at different temperatures.}
\end{figure}

\begin{figure}[h!]
	\centering
	\includegraphics[width=0.49\linewidth]{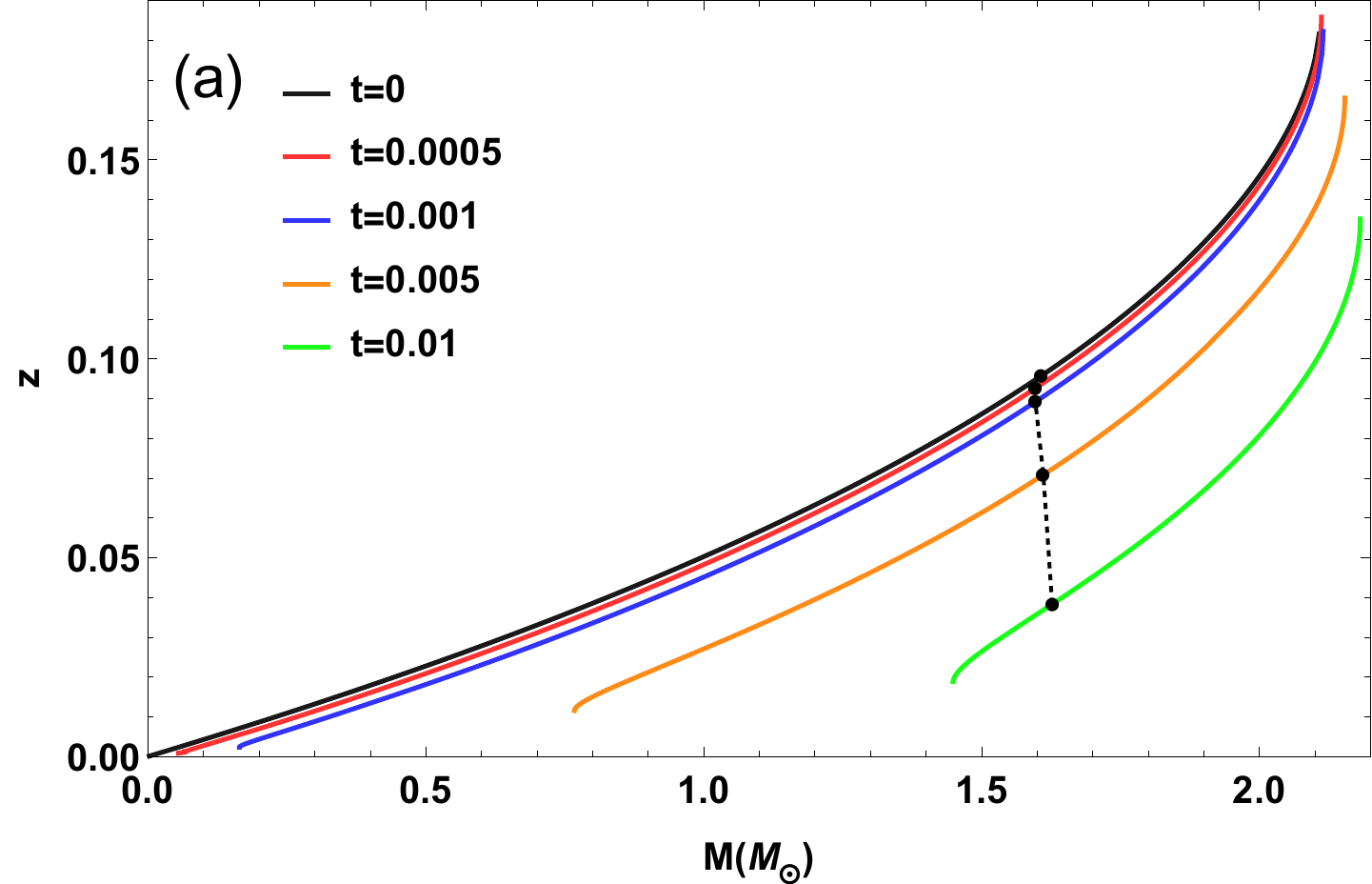}
	\includegraphics[width=0.49\linewidth]{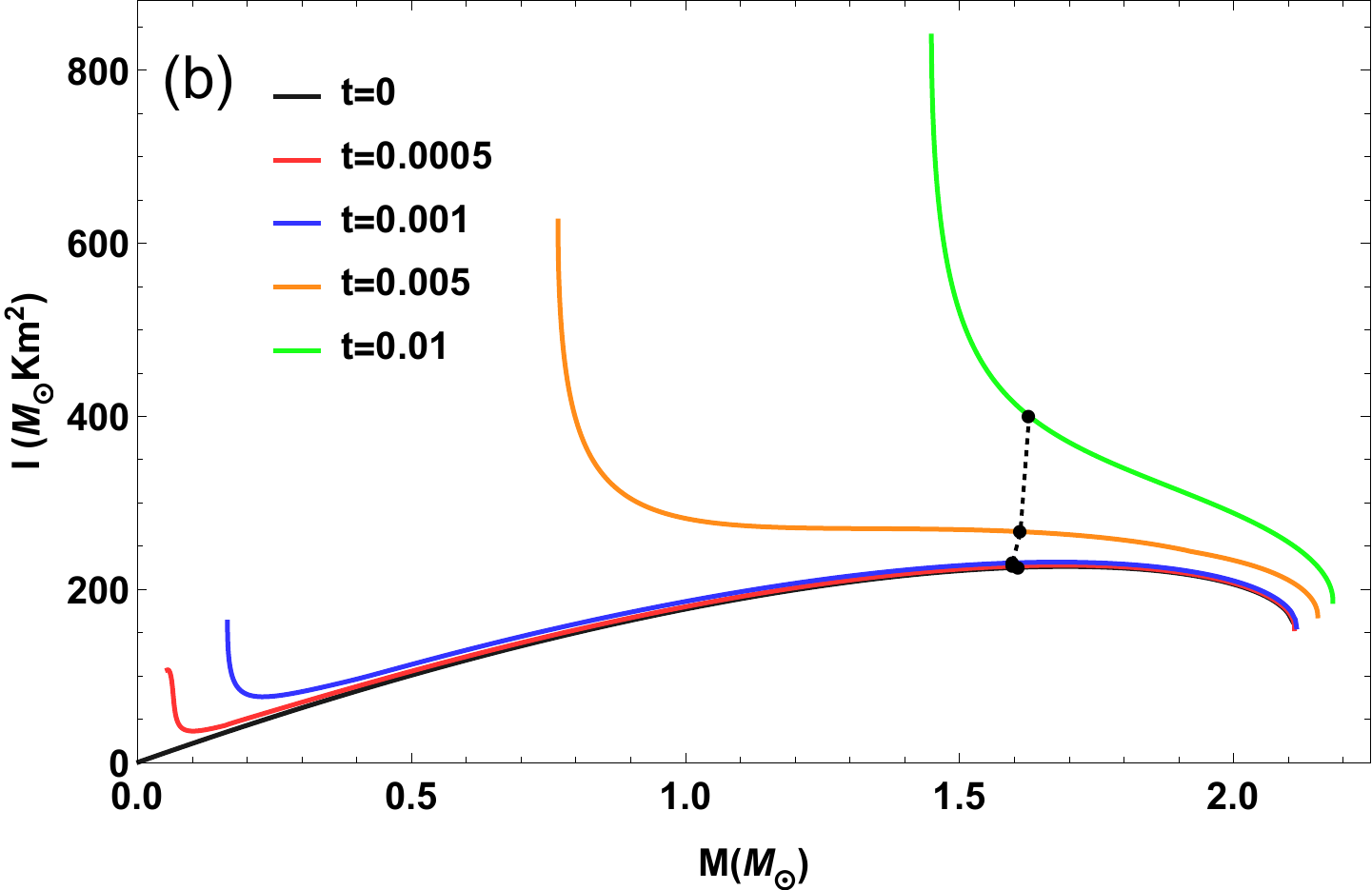}
	\caption{\label{F5}Gravitational redshift (a) and moment of inertia as functions of the mass of the star (b) for several values of the temperature.}
\end{figure}

Apart from increasing the mass and reducing the number of stable stars, the temperature increases the radius of the stars (Fig.~\ref{F4}a). As a consequence,  the compactness of hot BECS decreases with $T$, indicating that the hotter stars are less denser (Fig.~\ref{F4}b). An interesting feature of having BECS at different temperatures is that, for a fixed baryon mass, one can compare how the macroscopic properties of the star evolve as it gets colder and oldest. We did so for a star with a baryon mass of $1.7 M_{\odot}$ (composed by around $10^{57}$ bosons). The position of this star is marked with black dots on Figs.~\ref{F4} and \ref{F5}. Especially from Fig.~\ref{F4}b, we can see how the star becomes denser as its temperature decreases. This is caused by the reduction in the radius since the gravitational mass remains almost constant (Fig.~\ref{F4}a), indicating that the star contracts as it cools (its pressure decreases with $T$).

Finally, in Fig.~\ref{F5} we show the change in temperature of the redshift (a) and the moment of inertia (b). The redshift increases with decreasing temperature, because the more compact a star is, the more relativistic it gets.
As we can expect following the same reasoning, the moment of inertia increases with temperature, since it is proportional to the radius of the star.

\subsection{Magnetized Bose-Einstein condensate stars at finite temperature}\label{se45}

In this section we will analyze the interplay of temperature and magnetic field on the macroscopic properties of BEC solving Eqs.~(\ref{gTOV}) with the EoS Eqs.~(\ref{EoSR}) at $B=10^{17}$~G (Fig.~\ref{F2}b). This value of $B$ guarantees that the order of the thermal effects is comparable to that of the magnetic field.

\begin{figure}[h!]
	\centering
	\includegraphics[width=0.49\linewidth]{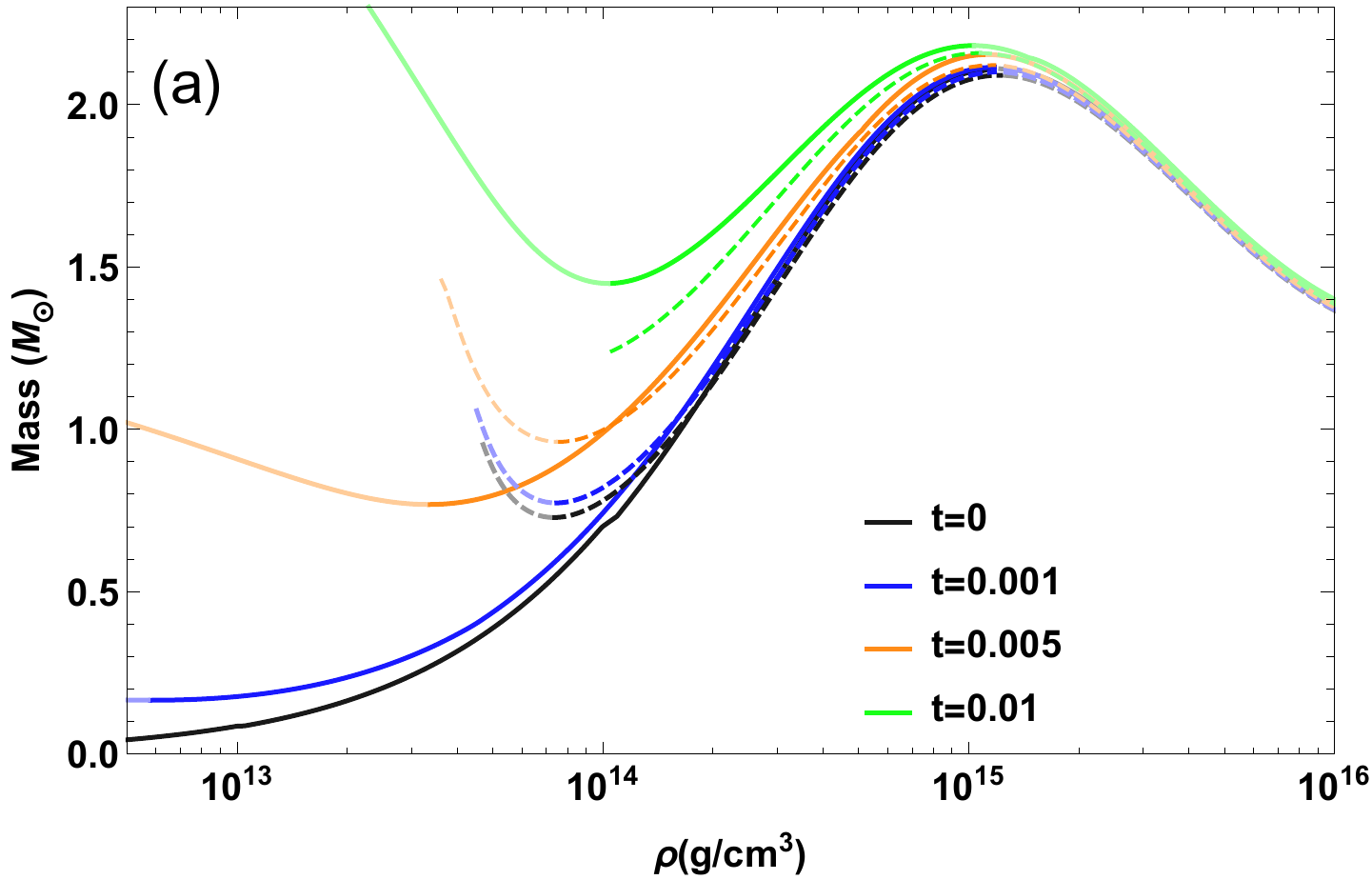}
	\includegraphics[width=0.49\linewidth]{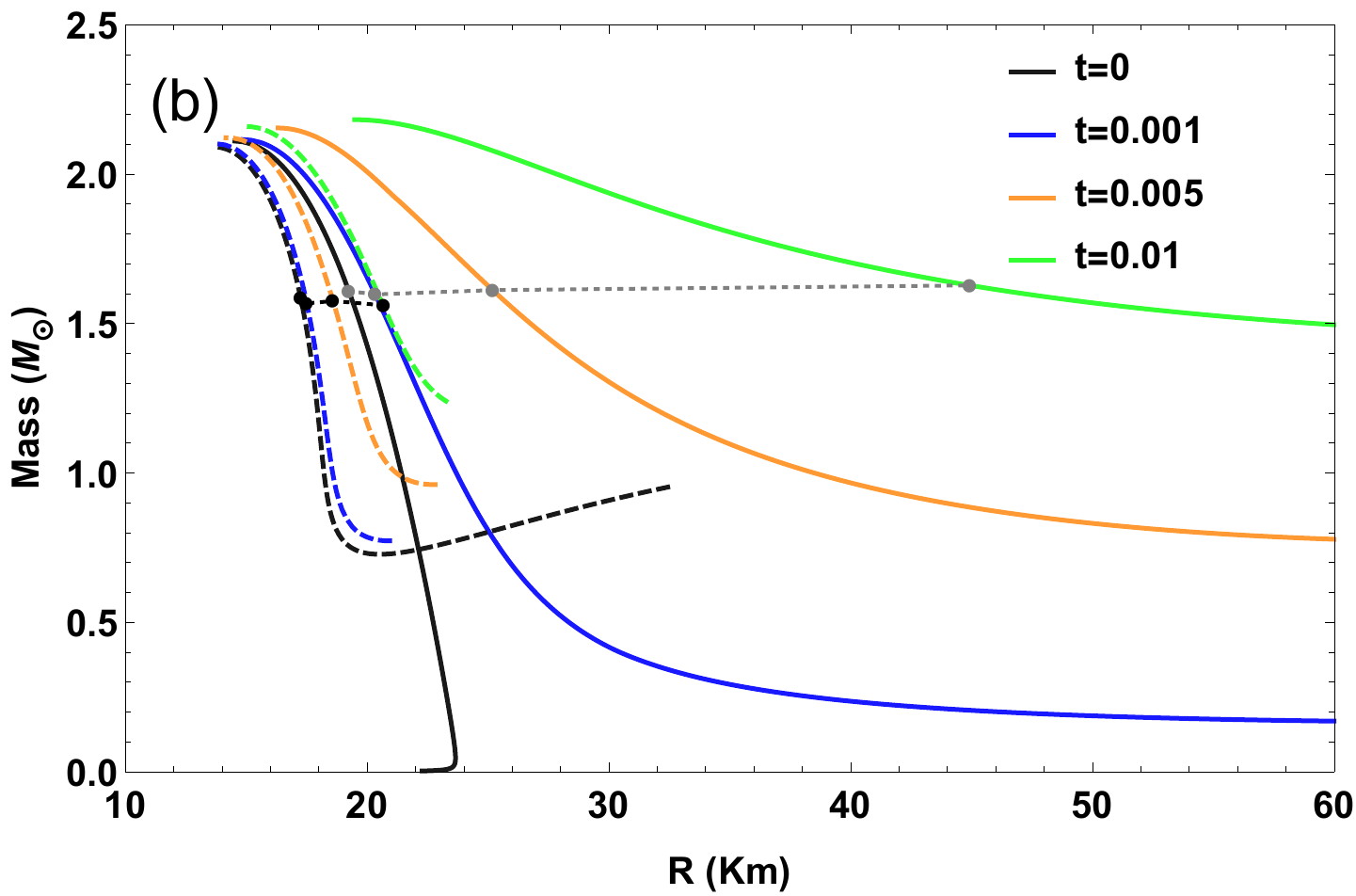}
	\caption{\label{F7} Mass-central density (a) and mass-mean radius (b) curves for the magnetized (dashed) and non-magnetized (solid) BECS at several temperatures. For all magnetized stars $B=10^{17}$~G.}
\end{figure}

For a fixed temperature, having a constant magnetic field reduces the number of stable star configurations (Fig.~\ref{F7}a). This reduction is not only due to the instability in $P_{\parallel}$ at low densities but to the change in the monotony of the $M(\rho)$ curve in this region. All the mass-central density curves of magnetized BECS are lower-bounded by the density at which $P_{\parallel}$ becomes zero, but the ones corresponding to $t=0$, $t=0.001$ and $t=0.005$ has a minimum in the low-density region that reduces, even more, the number of stable stars.

As for $B=0$, the maximum mass remains dominated by the characteristics of the bosons, and it is barely affected by the magnetic field, while the minimum mass of stable stars increases for $t=0$, $t=0.001$ and $t=0.005$ and decreases for $t=0.01$ (Table \textbf{II}). Note that for the smaller temperatures, the stars at the lower allowed central densities are slightly heavier than the non-magnetized stars that correspond to the same temperature. This non-monotonous behavior of $M$ with $B$ and $T$ may be a signature of the competition of the thermal and the magnetic effects.

The magnetic field drastically reduces the size of the stars. However, from Fig.~\ref{F7}b we can appreciate that for a high enough temperature, a hot magnetized BECS can attain masses and radii above those corresponding to the $B=0$, $T=0$ case (see the green dashed curve). In this sense, the thermal and magnetic effects do oppose each other.

\begin{table}[]
	\begin{tabular}{ccccccccc}
		\cline{2-9}
		\multicolumn{1}{c|}{\multirow{2}{*}{\textbf{}}} & \multicolumn{2}{c|}{$\rho_{min} (10^{12}g/cm^{3})$}                              & \multicolumn{2}{c|}{$\rho_{max}(10^{15}g/cm^3$)}                                & \multicolumn{2}{c|}{$M_{min}(M)$}                                              & \multicolumn{2}{c|}{$M_{max}(M$)}                                              \\ \cline{2-9}
		\multicolumn{1}{c|}{}                           & \multicolumn{1}{c|}{\textbf{B=0}} & \multicolumn{1}{c|}{\textbf{B=$10^{17}$G}} & \multicolumn{1}{c|}{\textbf{B=0}} & \multicolumn{1}{c|}{\textbf{B=$10^{17}$G}} & \multicolumn{1}{c|}{\textbf{B=0}} & \multicolumn{1}{c|}{\textbf{B=$10^{17}$G}} & \multicolumn{1}{c|}{\textbf{B=0}} & \multicolumn{1}{c|}{\textbf{B=$10^{17}$G}} \\ \hline
		\multicolumn{1}{|c|}{\textbf{t=0}}              & \multicolumn{1}{c|}{-}            & \multicolumn{1}{c|}{73.7}                  & \multicolumn{1}{c|}{1.23}         & \multicolumn{1}{c|}{1.23}                  & \multicolumn{1}{c|}{-}            & \multicolumn{1}{c|}{0.72}                  & \multicolumn{1}{c|}{2.11}         & \multicolumn{1}{c|}{2.09}                  \\ \hline
		\multicolumn{1}{|c|}{\textbf{t=0.001}}          & \multicolumn{1}{c|}{5.80}         & \multicolumn{1}{c|}{74.9}                  & \multicolumn{1}{c|}{1.18}         & \multicolumn{1}{c|}{1.18}                  & \multicolumn{1}{c|}{0.16}         & \multicolumn{1}{c|}{0.77}                  & \multicolumn{1}{c|}{2.11}         & \multicolumn{1}{c|}{2.10}                  \\ \hline
		\multicolumn{1}{|c|}{\textbf{t=0.005}}          & \multicolumn{1}{c|}{33.6}         & \multicolumn{1}{c|}{75.5}                  & \multicolumn{1}{c|}{1.12}         & \multicolumn{1}{c|}{1.18}                  & \multicolumn{1}{c|}{0.77}         & \multicolumn{1}{c|}{0.96}                  & \multicolumn{1}{c|}{2.15}         & \multicolumn{1}{c|}{2.12}                  \\ \hline
		\multicolumn{1}{|c|}{\textbf{t=0.01}}           & \multicolumn{1}{c|}{103}          & \multicolumn{1}{c|}{106}                   & \multicolumn{1}{c|}{1.03}         & \multicolumn{1}{c|}{1.07}                  & \multicolumn{1}{c|}{1.45}         & \multicolumn{1}{c|}{1.24}                  & \multicolumn{1}{c|}{2.18}         & \multicolumn{1}{c|}{2.16}                  \\ \hline
		\multicolumn{1}{l}{}                            & \multicolumn{1}{l}{}              & \multicolumn{1}{l}{}                       & \multicolumn{1}{l}{}              & \multicolumn{1}{l}{}                       & \multicolumn{1}{l}{}              & \multicolumn{1}{l}{}                       & \multicolumn{1}{l}{}              & \multicolumn{1}{l}{}
	\end{tabular}\label{table2}
	\caption{Extreme values of the central mass density and the total mass of the stable stellar configurations at finite temperature and $B=10^{17}$~G.}
\end{table}

In Fig.~\ref{F7}b we consider the effect of the magnetic field on the evolution of a star of baryon mass of $1.7 M_{\odot}$ (black dots on the plots). Similarly to $B=0$, at $B=10^{17}$~G the star becomes denser as its temperature decreases since its gravitational mass remain almost constant while its radii diminish (Fig.~\ref{F7}a). The only difference is that the gravitational mass of the hot magnetized star is slightly less than the one of the $B=0$ case (gray dots on the plots). If we were to study the joint evolution of the temperature and the magnetic field as the star ages, we should consider a situation in which both magnitudes diminish. Given their almost opposite effects on the star's size and mass, it will be hard to anticipate whether the BECS will contract or expand since this will depend on the relative values of $T$ and $B$.

\begin{figure}[h!]
	\centering
	\includegraphics[width=0.49\linewidth]{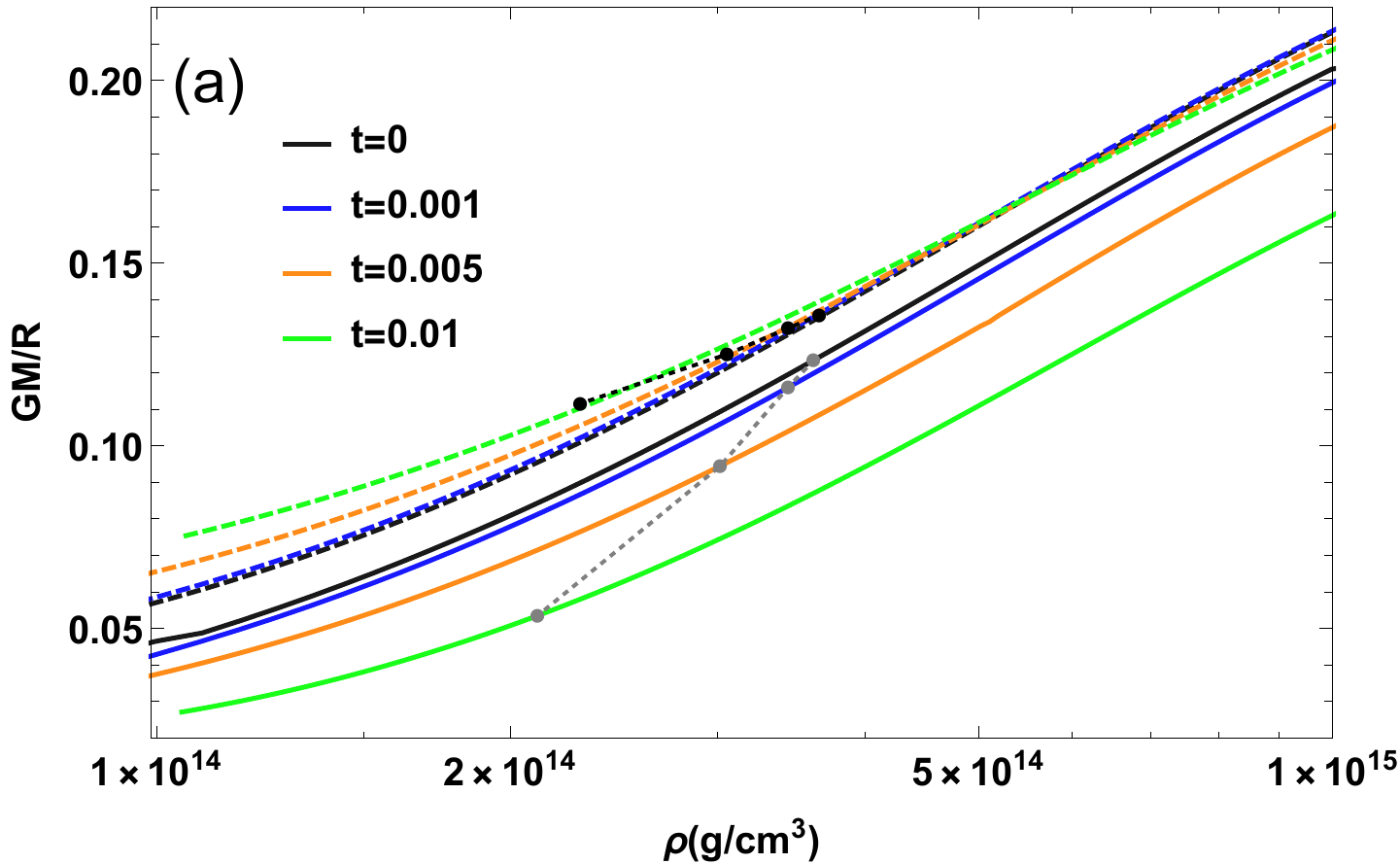}
	\includegraphics[width=0.49\linewidth]{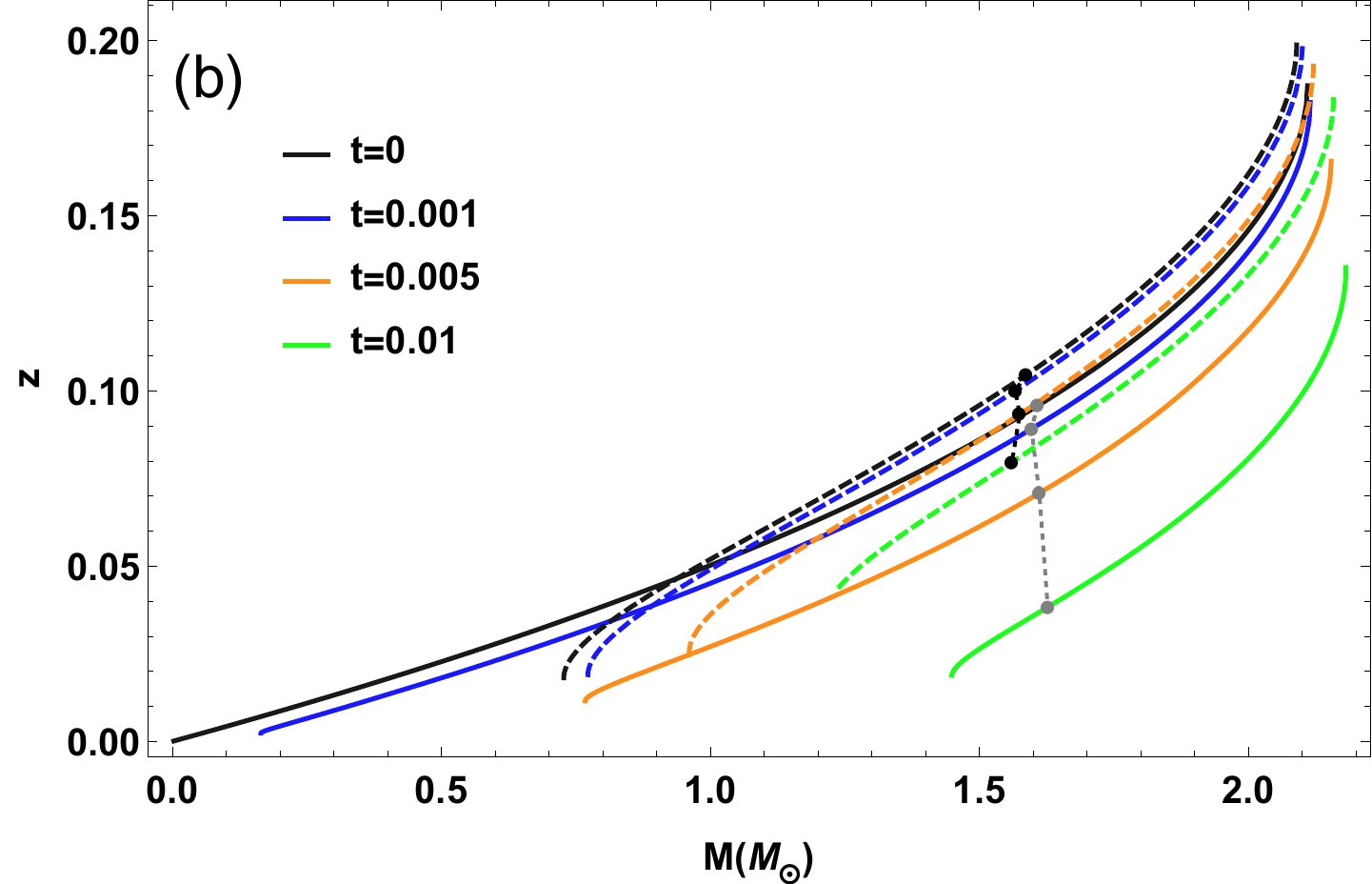}
	\caption{\label{F8} Compactness (a) and gravitational redshift (b) for the magnetized (dashed) and non-magnetized (solid) BECS at several temperatures.}
\end{figure}

Due to the reduction of the size of the stars caused by the magnetic field, the compactness of magnetized hot BECS is above that of the $B=0$, $T=0$ case for all the considered temperatures (Fig.~\ref{F8}a). The redshift also increases at a finite magnetic field, meaning that the star becomes more "relativistic" (Fig.~\ref{F8}b). However, for a fixed magnetic field, the compactness increases with temperature for low density, while at higher densities decreases with $T$. Also from Fig.~\ref{F8}b we see that as the star cools, its central density augments.

\begin{figure}[h!]
	\centering
	\includegraphics[width=0.315\linewidth]{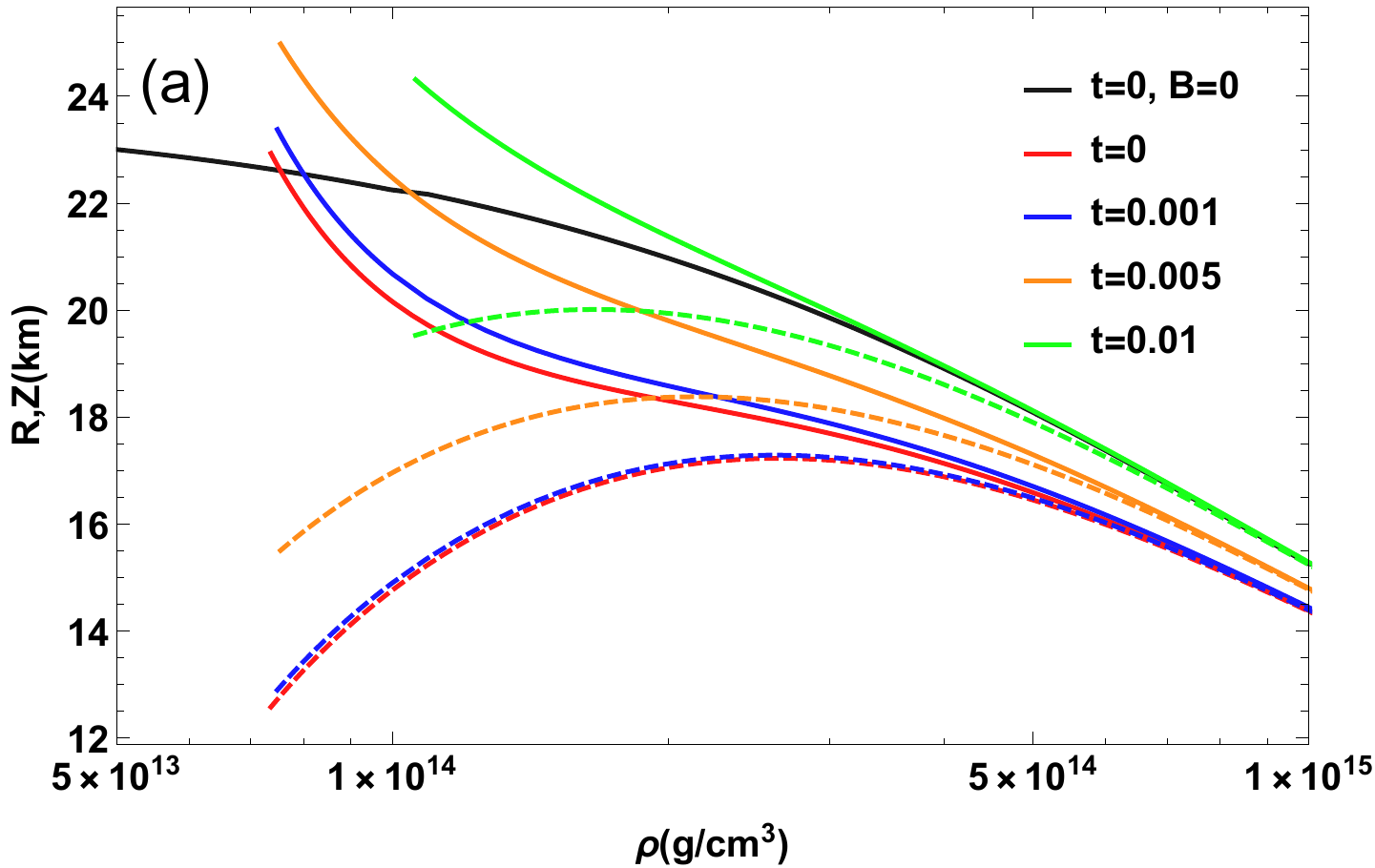}
	\includegraphics[width=0.31\linewidth]{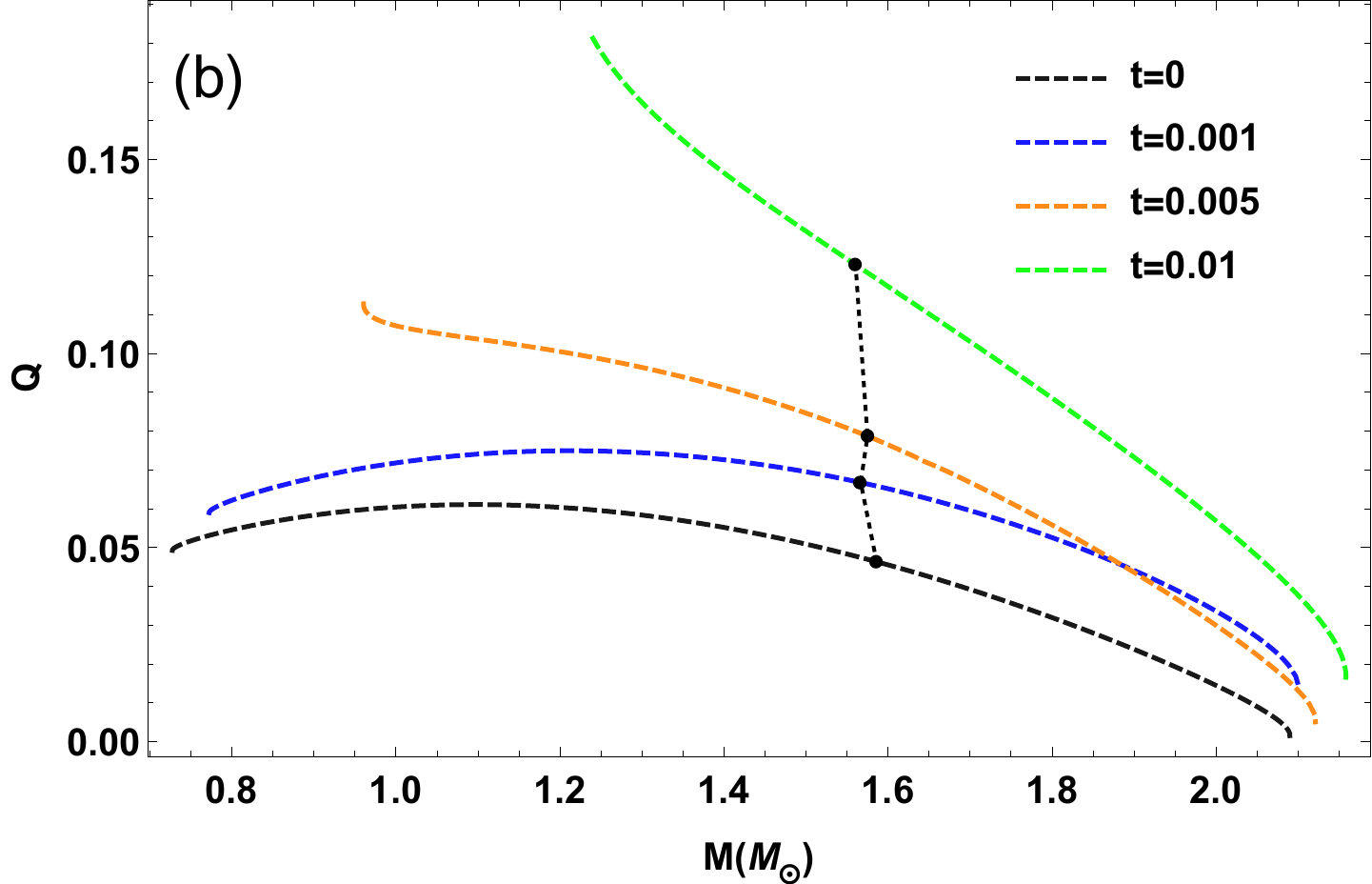}
	\includegraphics[width=0.295\linewidth]{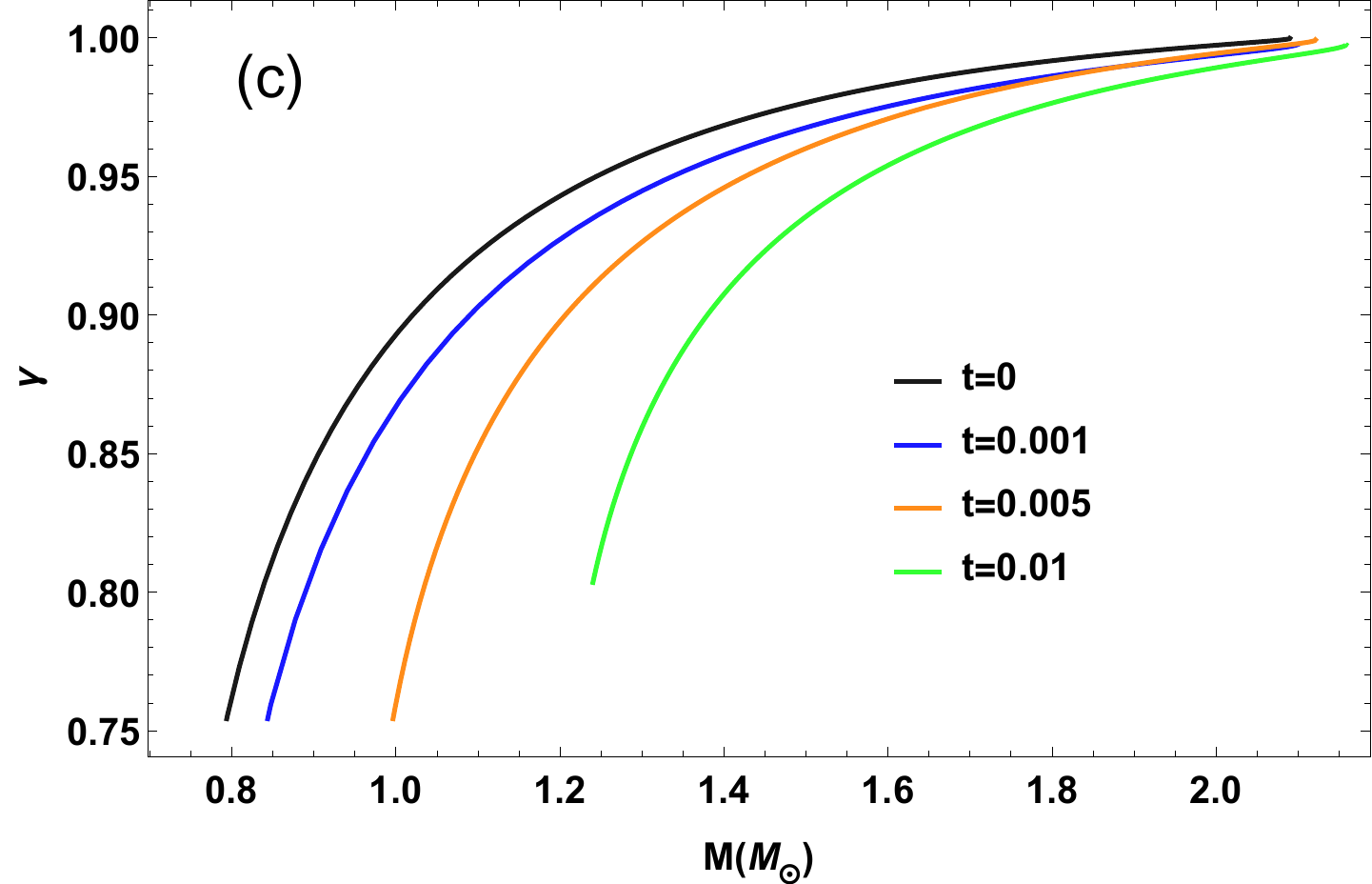}
	\caption{\label{F6} Equatorial (solid) and polar (dashed) radius as a function of density for $B=10^17$~G and several values of temperature (a). The mass quadrupolar moment as a function of the star mass (b). The parameter $\gamma$ as a function of the star mass (c).}
\end{figure}

To explore the effects of the temperature on the deformation of the magnetized BECS, in Fig. \ref{F6}a we show the equatorial (solid) and polar (dashed) radius $R$ and $Z$ as functions of the stars' central density for several values of the temperature. At $T=0, B \neq 0$, $Z<R$, hence, the star is an oblate object, and the difference between $R$ and $Z$ is higher at low densities. An increase in temperature increases $R$ and $Z$ without eliminating their difference. Thus, hot magnetized stars are still oblate and exhibit a non-zero mass quadrupolar moment that increases with $T$ (Fig.~\ref{F6}b). This happens because the hotter stars are more deformed than the colder ones (their $\gamma$ is farther from one, Fig.~\ref{F6}c).
Note that $Q$ depends directly on $M$ and $\gamma$ (Eq.~(\ref{Q})). The values of $Q$ for the more massive stars are close to zero since for those stars $\gamma\simeq 1$. As the gravitational mass decreases, the behavior of $Q$ changes according to temperature. For the smallest values of $T$, at which the magnetic effects dominate, $Q$ attains a maximum value for an intermediate mass. This also happens for magnetized strange stars \cite{Samantha}, and it is connected with the fact that at high mass $\gamma$ goes to $1$ ($P_{\parallel_0}\simeq P_{\perp_0}$), and at lower mass $\gamma$ decreases ($P_{\parallel_0}\rightarrow0$). As $T$ increases, the sequence of stable stars is reduced, the maximum disappears, and the lighter stars become the ones with the highest value of $Q$. Regarding gravitational waves, the behavior of $Q$ with $T$ implies that the temperature may enhance their emission for the stars of small and intermediate masses.

\section{Concluding remarks}\label{sec5}

We obtained the EoS and the macroscopic properties of relativistic BECS at finite temperatures with and without a magnetic field. Such stars consist of a condensed gas of interacting neutral vector bosons coupled to a uniform and constant magnetic field. We assumed for the bosons a two-body contact interaction independent of the temperature and the magnetic field, while for the thermal part of the EoS we used the one-loop thermodynamic potential of a gas of free vector bosons at a finite temperature under the action of an external magnetic field. To obtain the macroscopic properties we used the $\gamma$-structure equations since they properly account for the anisotropy in the pressures caused by the magnetic field.

Our study was restricted to temperatures two orders below the boson mass. At $T>0.01 m$, the Bose-Einstein condensation is achieved at densities far beyond the central densities of stable stars. Thus, going above this limit does not preserve the nature of BECS. Since the antiparticles' density starts to be non-negligible at  $T\geq m$, pair production has no relevance in our model. Finally, the values of $T$ and $B$ used in the plots were selected such that the thermal and magnetic energy is about the same order. This makes it possible to analyze the joint effects of $T$ and $B$ in the physics of the star.

At zero magnetic field, the main effect of the temperature in the EoS of BECS is to increase the pressure at low densities. It reflects later as an increase in the masses and radius of the stars. Thus hot BECS are bigger and heavier than their zero-temperature counterparts. These effects are more relevant for the stars with lower central densities and higher temperatures, while the maximum mass of the model is almost unchanged and remains mainly dominated by the microscopic properties of the bosons. The other relevant and unexpected feature related to the pure thermal effects is the existence of a lower bound on the central density of stable stars. This lower allowable density increases with $T$, such that the higher the temperature, the less the number of stable BECS.

The magnetic field presence reduces, even more, the number of stable stars and provokes the BECS to axially deform in a manner that affects their size and other macroscopic properties. In general, magnetized BECS at finite $T$ are smaller, denser, and more compact than non-magnetized hot stars. Although increasing $T$ augments the mass and size of the magnetized stars, a finite temperature is not enough to erase neither the anisotropy in the pressure nor the instability caused by the magnetic field at low densities. Moreover, increasing the temperature favors the deformation and augments the quadrupolar moment of the stars of small and intermediate masses, apparently enhancing gravitational waves' emission. Therefore, if at first sight it might seem that a finite temperature will cancel the magnetic field effects in BECS, it is clear from the results of our study that the interaction of $T$ and $B$ adds new and non-trivial features to the physics of these stars.

\section{Acknowledgments}
The work of G.Q.A, L.C.S.G and H.P.R. was supported by project No. NA211LH500-002 of AENTA-CITMA, Cuba. A.P.M acknowledge the support of the Agencia Estatal de Investigaci\'on through the grant PID2019-107778GB-100 from Junta de Castilla y Le\'on, Spanish Consolider MultiDark FPA2017-90566-REDC and PHAROS COST Actions MP1304 and CA16214.

\end{document}